\documentstyle[prc,aps,wrapft,epsf,psfig]{revtex}

\def\Journal#1#2#3#4{{#1} {\bf #2}, #3 (#4)}

\def\NP{{\em Nucl. Phys.}}

\def\PLB{{\em Phys. Lett.}  B}
\def\PRL{\em Phys. Rev. Lett.}
\def\PRD{{\em Phys. Rev.} D}
\def\PRC{{\em Phys. Rev.} C}
\def\ZPA{{\em Z. Phys.} A}
\def\ZPC{{\em Z. Phys.} C}
\def\PR{{\em Phys. Rep.}}
\def\CPC{{\em Comp. Phys. Comm.}}

\def\ARPS{{\em Annu. Rev. Part. Sci.}}
\def\ANP{{\em Annals Phys.}}
\def\PPNP{{\em Prog. Part. Nucl. Phys.}}
\def\HIP{{\em Heavy Ion Phys}}
\def\PTPS{{\em Prog. Theor. Phys. Suppl.}}
\newcommand{\comment}[1]{}

\newcommand{\AmS}{{\protect\the\textfont2
  A\kern-.1667em\lower.5ex\hbox{M}\kern-.125emS}}
\newcommand{\srt}{\mbox{$\sqrt{s}$}}
\def\GeV{{\rm GeV}}
\def\AGeV{{\rm A GeV}}
\def\GeVc{{\rm GeV/c}}
\def\AGeVc{{\rm A GeV/c}}

\def\fm{{\rm fm}}
\newcommand{\Dl}{\mbox{$\Delta(1232)$}}
\newcommand{\Ds}{\mbox{$\Delta^*$}}
\newcommand{\Ns}{\mbox{$N^*$}}
\newcommand{\Mx}{\mbox{${\cal M}$}}                    
\newcommand{\bold}[1]{\mbox{\boldmath $#1$}}    
\newcommand{\pp}{\bold p}
\newcommand{\plab}{\mbox{$p_{\rm lab}$}}
\newcommand{\Kbar}{\mbox{${\bar K}$}}
\newcommand{\Km}{\mbox{$K^-$}}

\newcommand{\sigtot}{\mbox{$\sigma_{tot}(s)$}}
\newcommand{\sigel}{\mbox{$\sigma_{el}(s)$}}
\newcommand{\sigtR}{\mbox{$\sigma_{\mbox{t-R}}(s)$}}
\newcommand{\sigtS}{\mbox{$\sigma_{\mbox{t-S}}(s)$}}
\newcommand{\sigbw}{\mbox{$\sigma_{BW}(s)$}}
\newcommand{\sigsS}{\mbox{$\sigma_{\mbox{s-S}}(s)$}}
\newcommand{\sigpiY}{\mbox{$\sigma_{\pi Y}(s)$}}

\newcommand{\sigch}{\mbox{$\sigma_{{\rm ch}}(s)$}}

\newcommand{\etal}{{\it et al.}}
\newcommand{\gsim}{\mbox{\raisebox{-0.6ex}{$\stackrel{>}{\sim}$}}\:}
\newcommand{\lsim}{\mbox{\raisebox{-0.6ex}{$\stackrel{<}{\sim}$}}\:}

\def\FIGnntotal{%
\begin{figure}
\centerline{\hbox{\psfig{figure=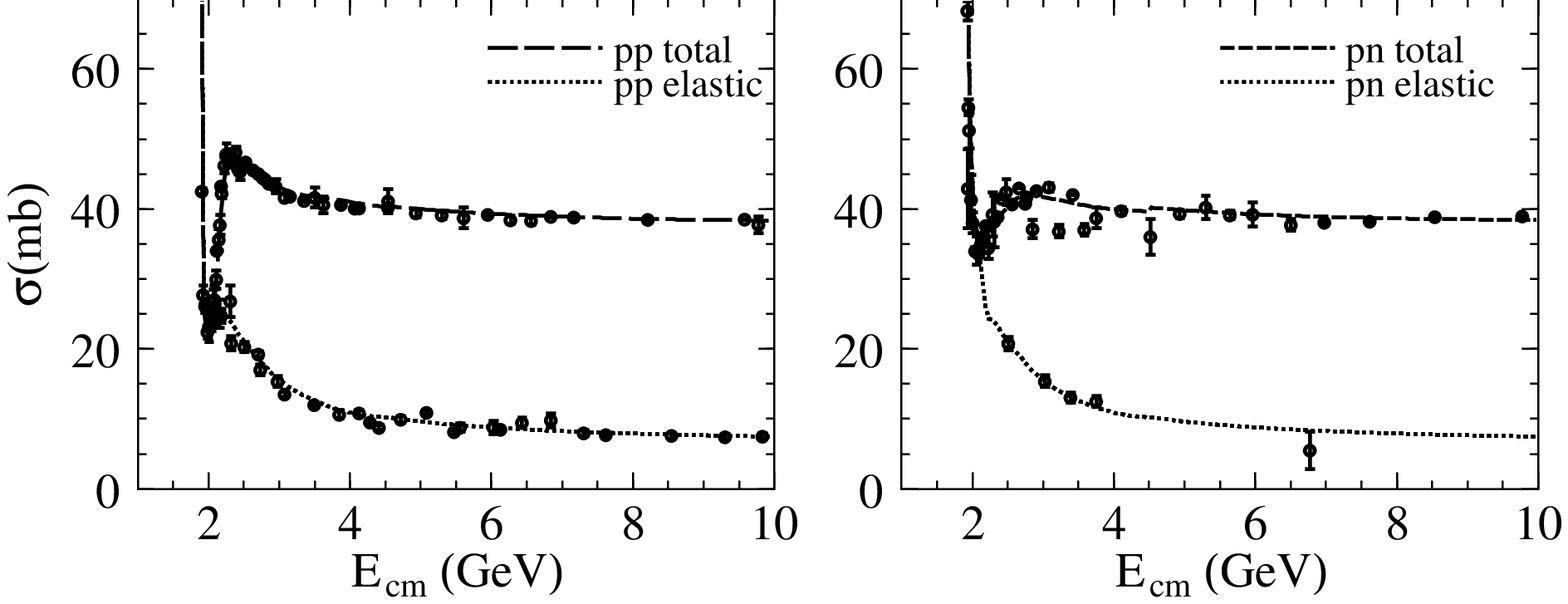,height=7.5cm}}}
\caption[]
      {
  The fitted total and elastic $pp$ and $pn$ cross sections
  which are used in the code
   together with measured data
  taken from particle data group~\cite{PDG96}.
  CERN-HERA and COMPAS group parameterization
   are used at high energy~\cite{PDG96}.
     }
   \label{fig:nntotal}
\end{figure}
}

\def\FIGnninel{%
\begin{figure}
\centerline{\hbox{\psfig{figure=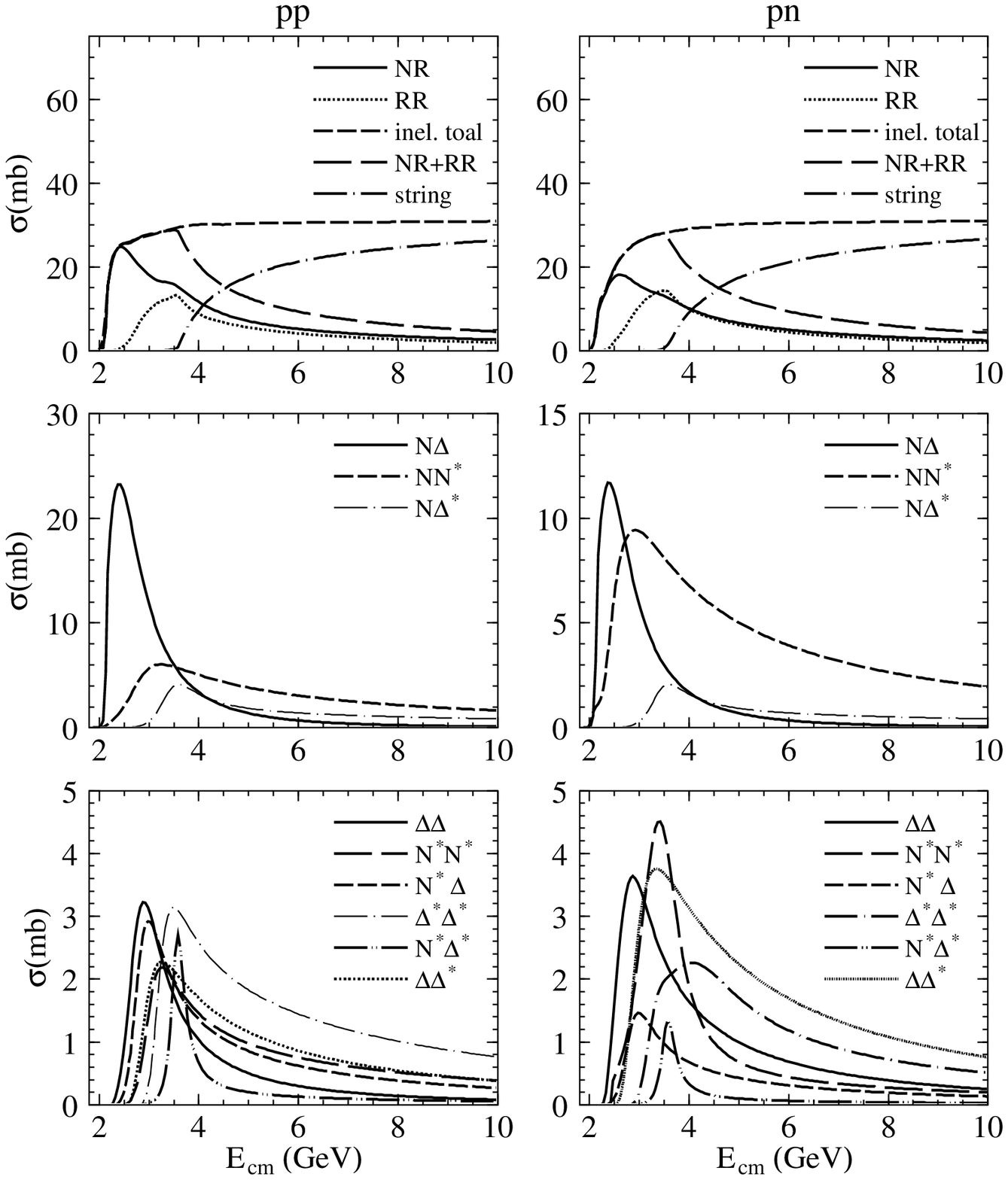,height=15cm}}}
\caption[]
      {
  The resonance production cross sections
  for $pp$ (left panels) and $pn$ (right panels) reactions
  as functions of c.m. energies.
  In the upper panels, the total one-resonance (NR),
  double-resonance (RR), total resonance (NR+RR), total inelastic
  cross section and string formation cross sections are shown.
  In the middle and the lower panels,
   each one-resonance production branches and
   double-resonance production branches are plotted.
     }
   \label{fig:nninel}
\end{figure}
}

\def\FIGnnex{%
\begin{figure}
\centerline{\hbox{\psfig{figure=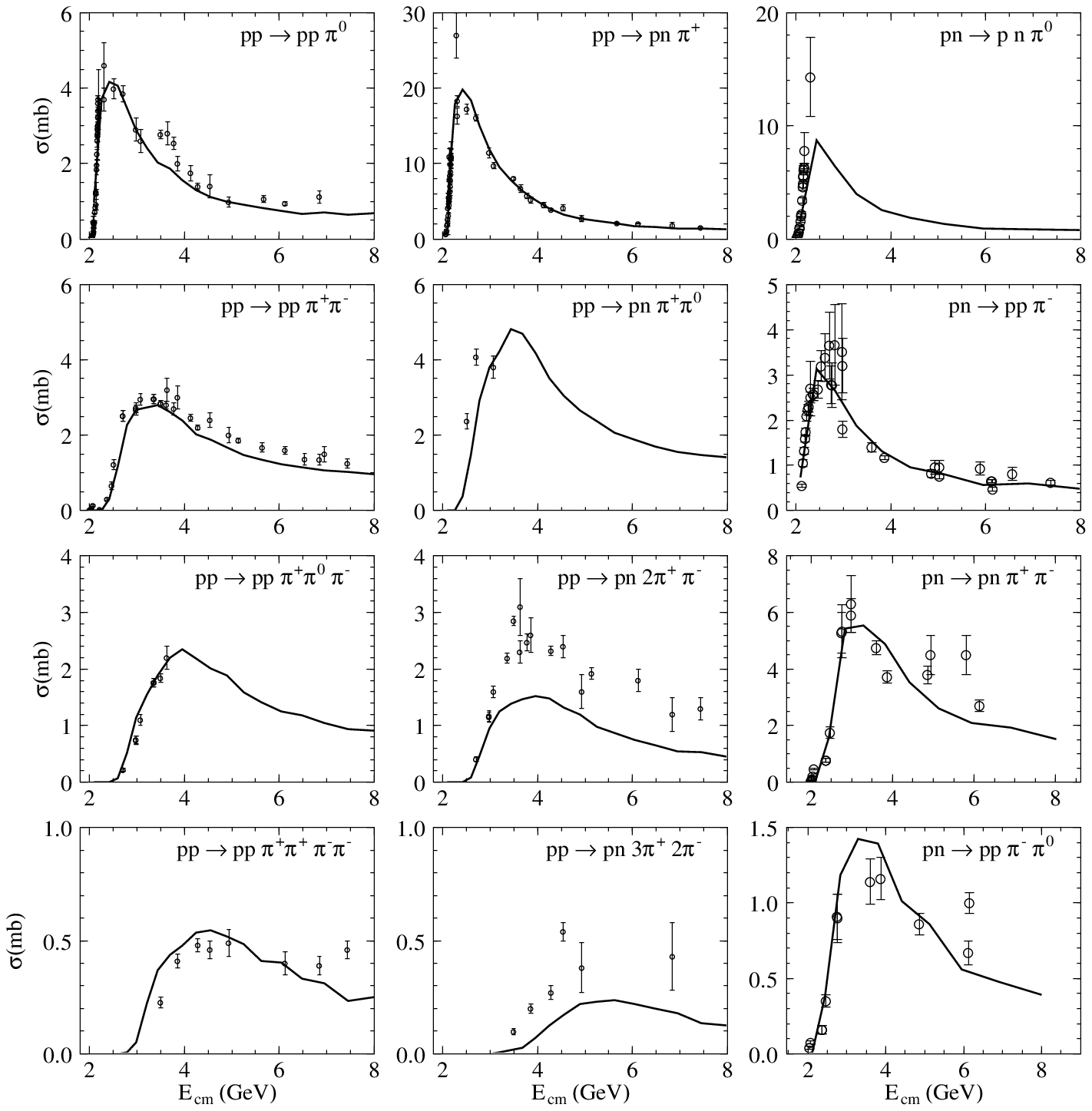,height=15cm}}}
\caption[]
      {
  The energy dependence of the exclusive pion production cross sections
  for proton-proton and proton-neutron interactions as a function of c.m. energy.
  Solid lines are the results obtained from the code.
  Data from Ref.~\cite{CernHera}.
     }
   \label{fig:nnex}
\end{figure}
}

\def\FIGdetbal{%
\begin{figure}
\centerline{\hbox{\psfig{figure=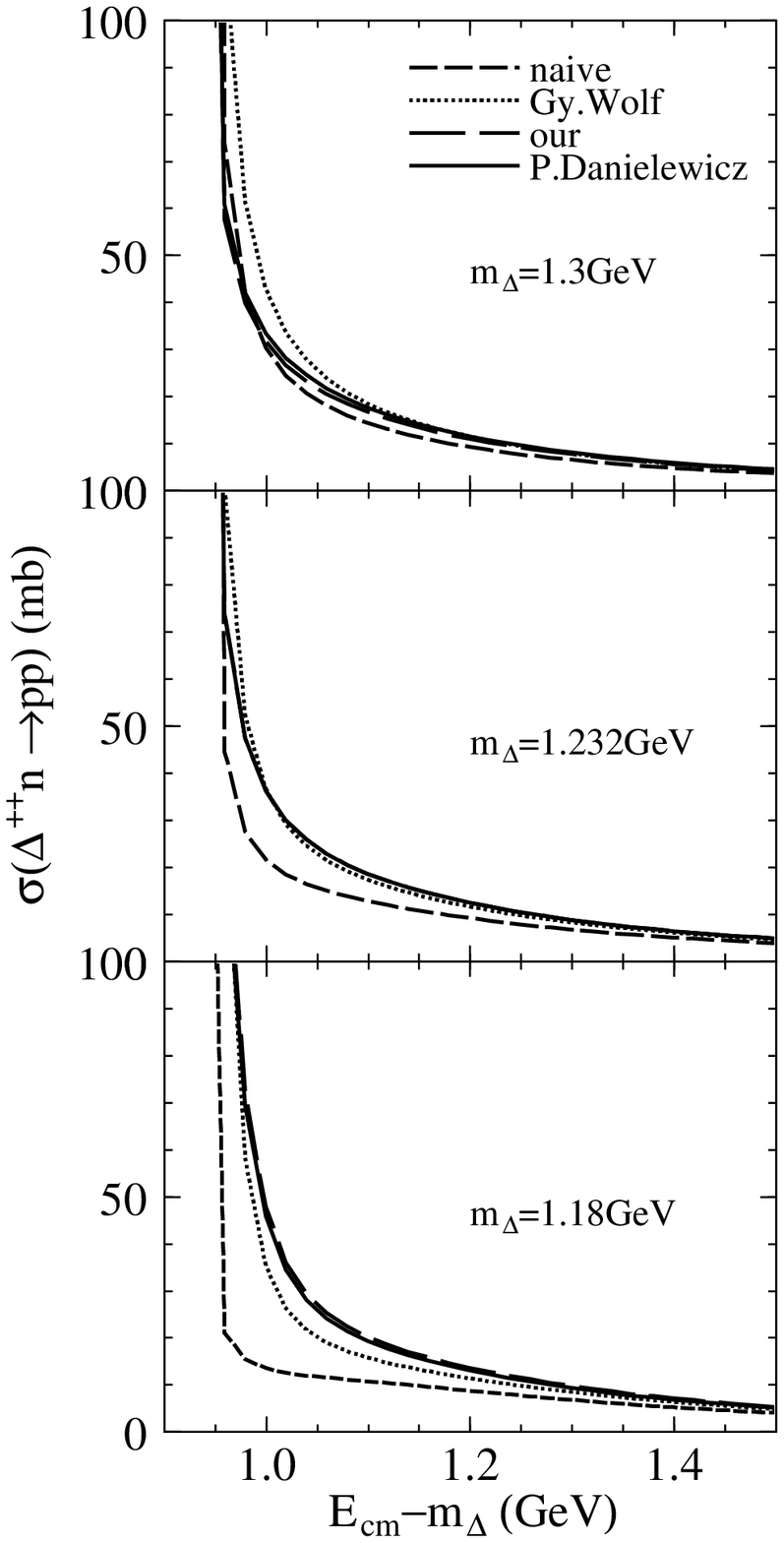,height=10cm}}}
\caption[]
      {
  The cross section for $\Delta^{++}n\to pp$
  calculated using the different descriptions for the
  detailed balance as a function of $\srt-M_{\Delta}$,
   where $M_{\Delta}$ is the mass of ingoing $\Delta$.
   The short-dashed lines correspond to the results of the formula which
   does not take the $\Delta$ width into account.
The results of the cross sections obtained using the formula in
 Ref.~\cite{Wolf2} and \cite{detbal1} are shown by
  dotted and full lines respectively.
 The long-dashed lines correspond to the results using
  Eq.(\ref{eq:detbaldelta}).
   
     }
   \label{fig:detbal}
\end{figure}
}

\def\FIGxpin{%
\begin{figure}[tb]
\begin{minipage}[t]{8.5cm}
\centerline{\psfig{figure=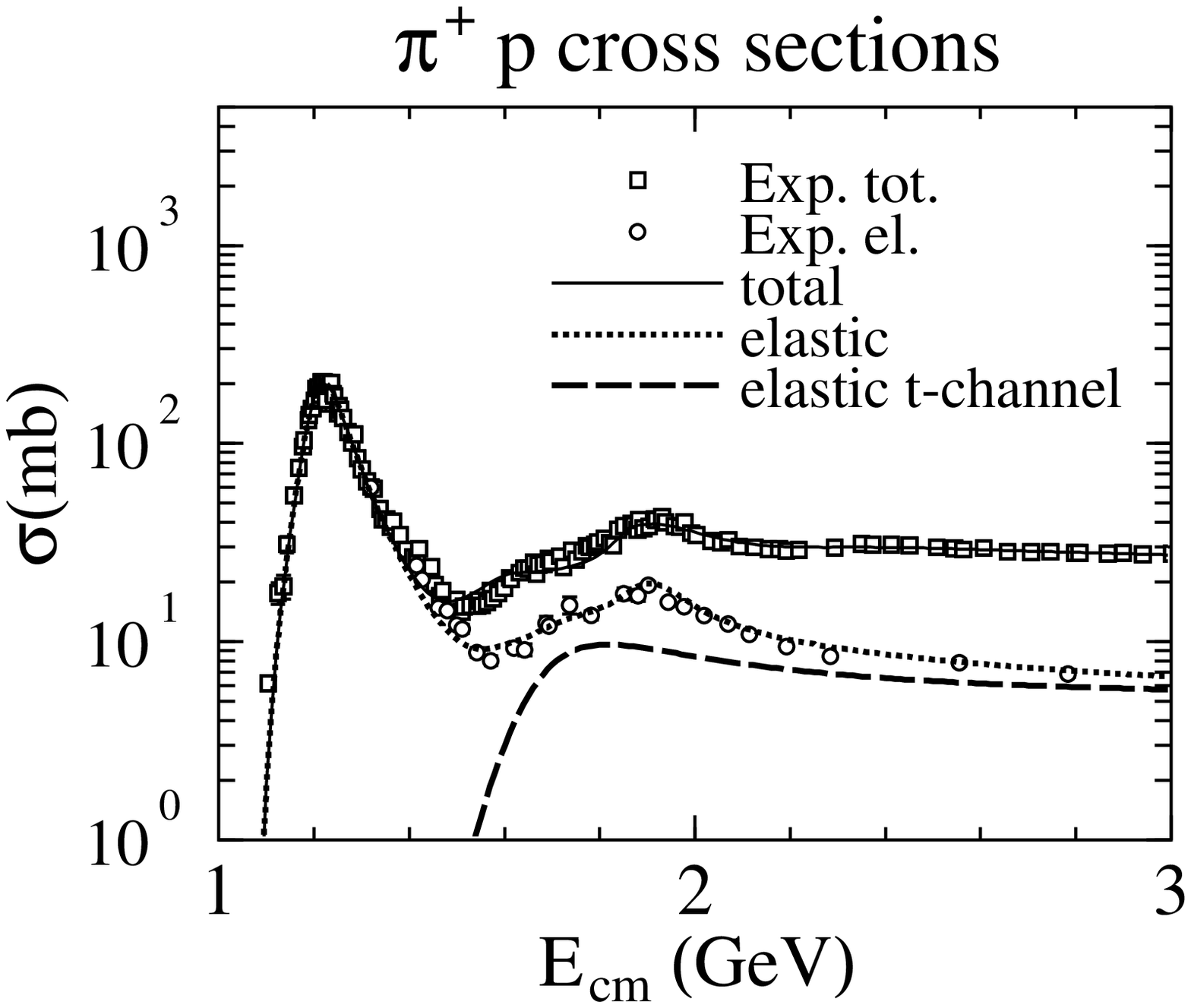,width=8.5cm}}
\caption{\label{fig:xpip}
Parameterization of the total and elastic $\pi^+p$ cross sections.
The data has been taken from \protect \cite{PDG96}.
Total and elastic cross sections are assumed to be the
sum of $s$-channel and $t$-channel resonance and/or string formation processes.
}
\end{minipage}
\hfill
\begin{minipage}[t]{8.5cm}
\centerline{\psfig{figure=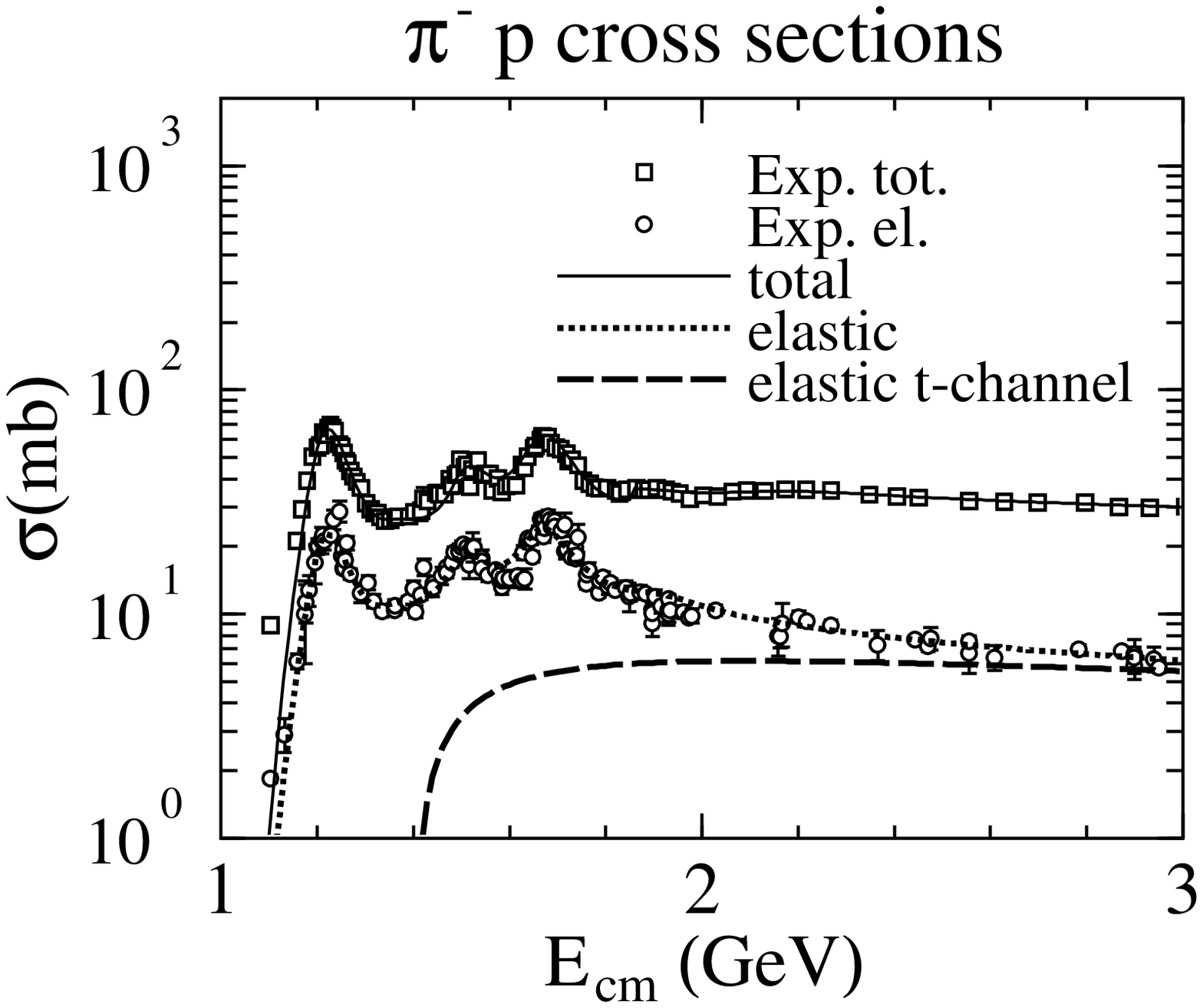,width=8.5cm}}
\caption{
   Parameterization of the total and elastic
   $\pi^-p$ cross sections.
   The data has been taken from \protect \cite{PDG96}.
\label{fig:xpin}
}
\end{minipage}
\end{figure}
}

\def\FIGxakn{%
\begin{figure}[tb]
\begin{minipage}[t]{8.5cm}
\centerline{\psfig{figure=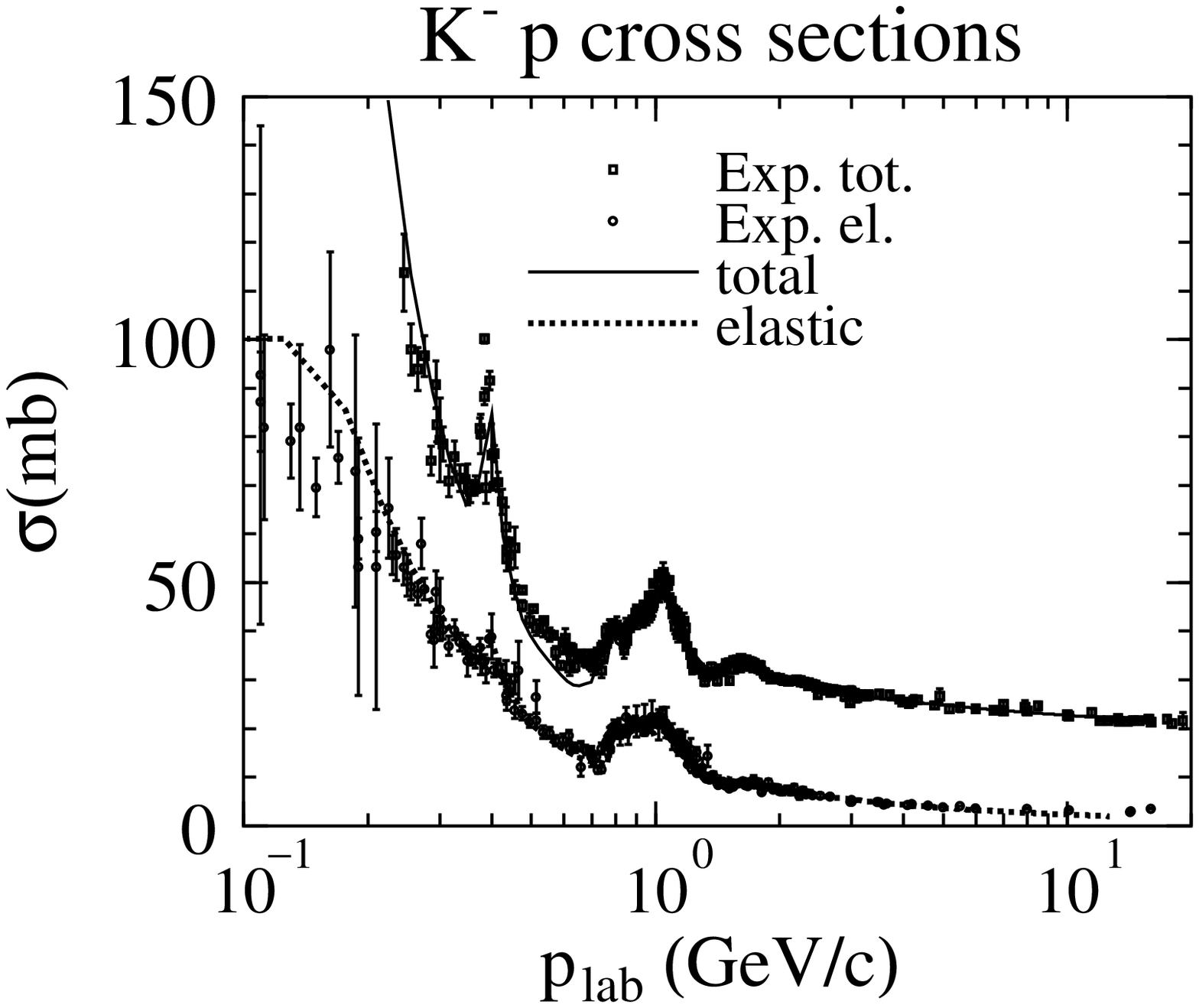,width=8.5cm}}
\caption{\label{fig:akn1}
  Parameterization of the total and elastic $K^-p$ cross sections.
  The data has been taken from \protect \cite{PDG96}.
  The sum of the $s$-channel resonance productions and $t$-channel elastic,
  $t$-channel charge exchange and $t$-channel hyperon production cross sections
  can  describe the data properly at low energies.}
\end{minipage}
\hfill
\begin{minipage}[t]{8.5cm}
\centerline{\psfig{figure=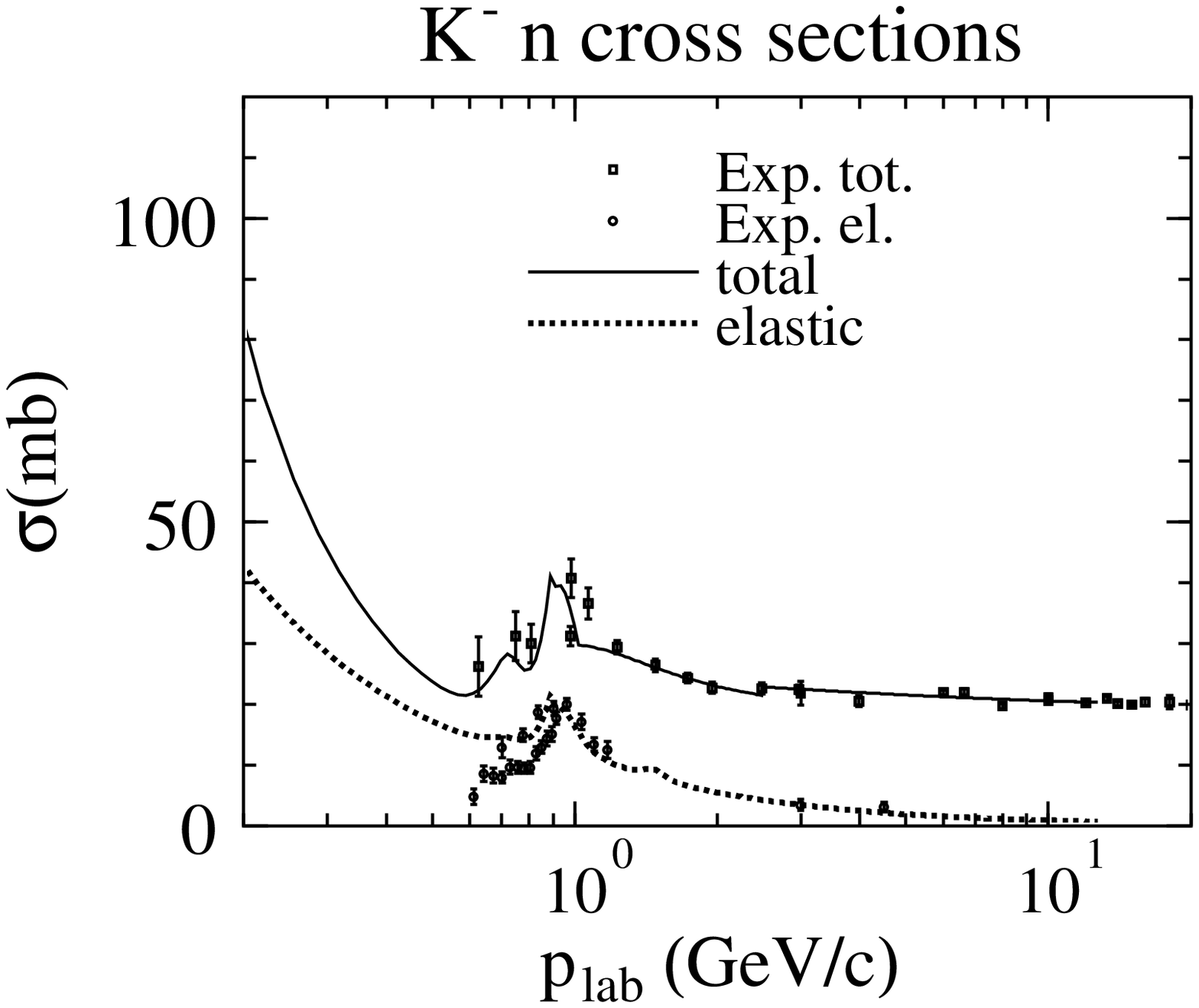,width=8.5cm}}
\caption{\label{fig:akn2}
Parameterization of the total and elastic $K^-n$ cross sections.
Parameterizations of low energy part of the cross section 
 where data is absent
 are addressed by Ref. \protect \cite{knscatt}.
The data has been taken from \protect \cite{PDG96}.
}
\end{minipage}
\end{figure}
}

\def\FIGxkny{%
\begin{figure}
\centerline{\hbox{\psfig{figure=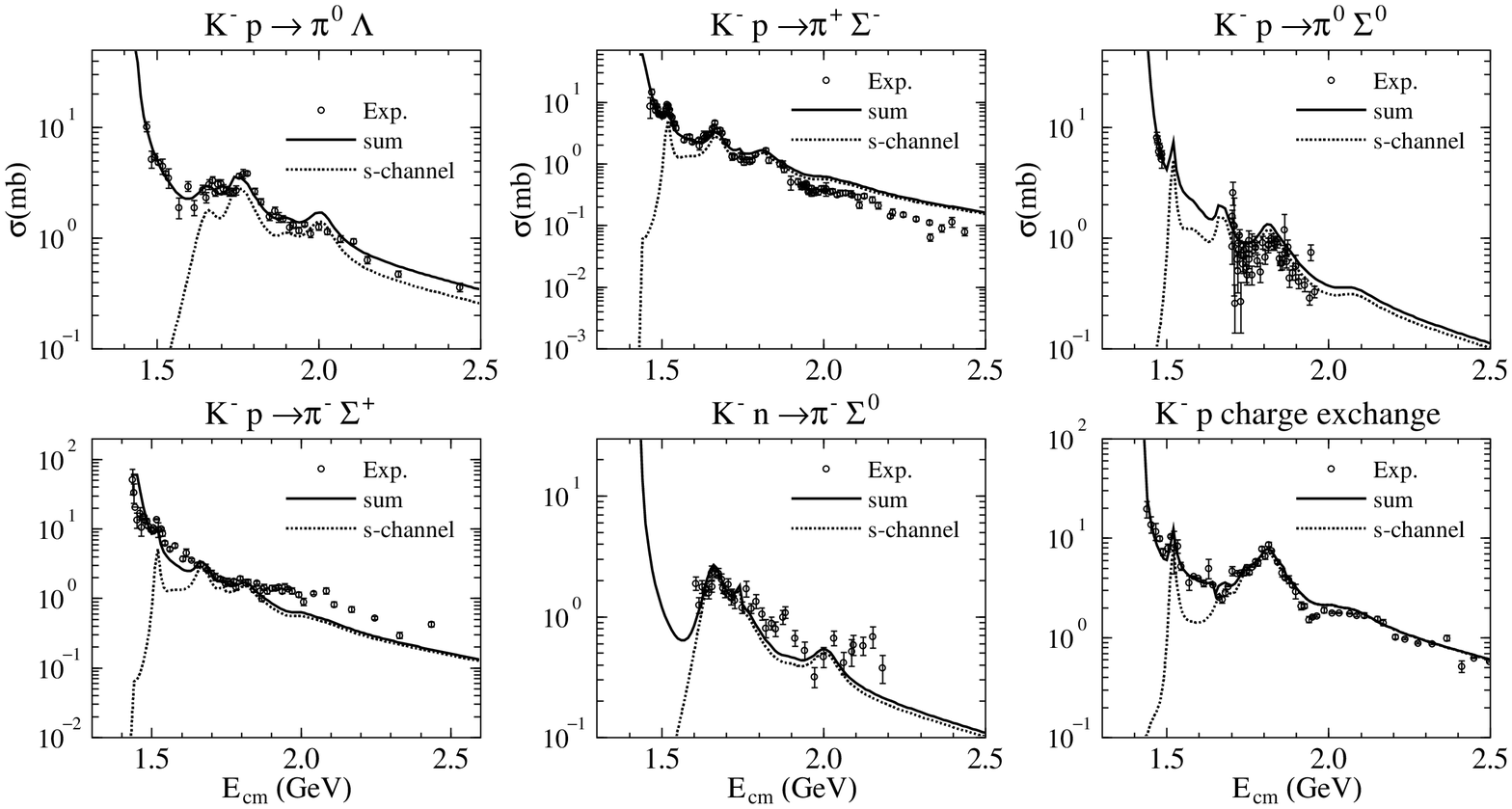,height=10cm}}}
\caption[]
      {
  The energy dependence of the exclusive hyperon production cross sections
  $\Km p\to \pi^0 \Lambda$,
  $\Km p\to \pi^- \Sigma^+$,
  $\Km p\to \pi^+ \Sigma^-$,
  $\Km n\to \pi^- \Sigma^0$,
  $\Km n\to \pi^0 \Sigma^0$,
  and charge exchange cross section $\Km p\to K^0 n$ used in the model
  are shown by the full lines
  together with $s$-channel (dotted) contributions.
 Circles are data from Ref.~\cite{HERA}.
     }
   \label{fig:xkny}
\end{figure}
}

\def\FIGxkaon{%
\begin{figure}
\centerline{\hbox{\psfig{figure=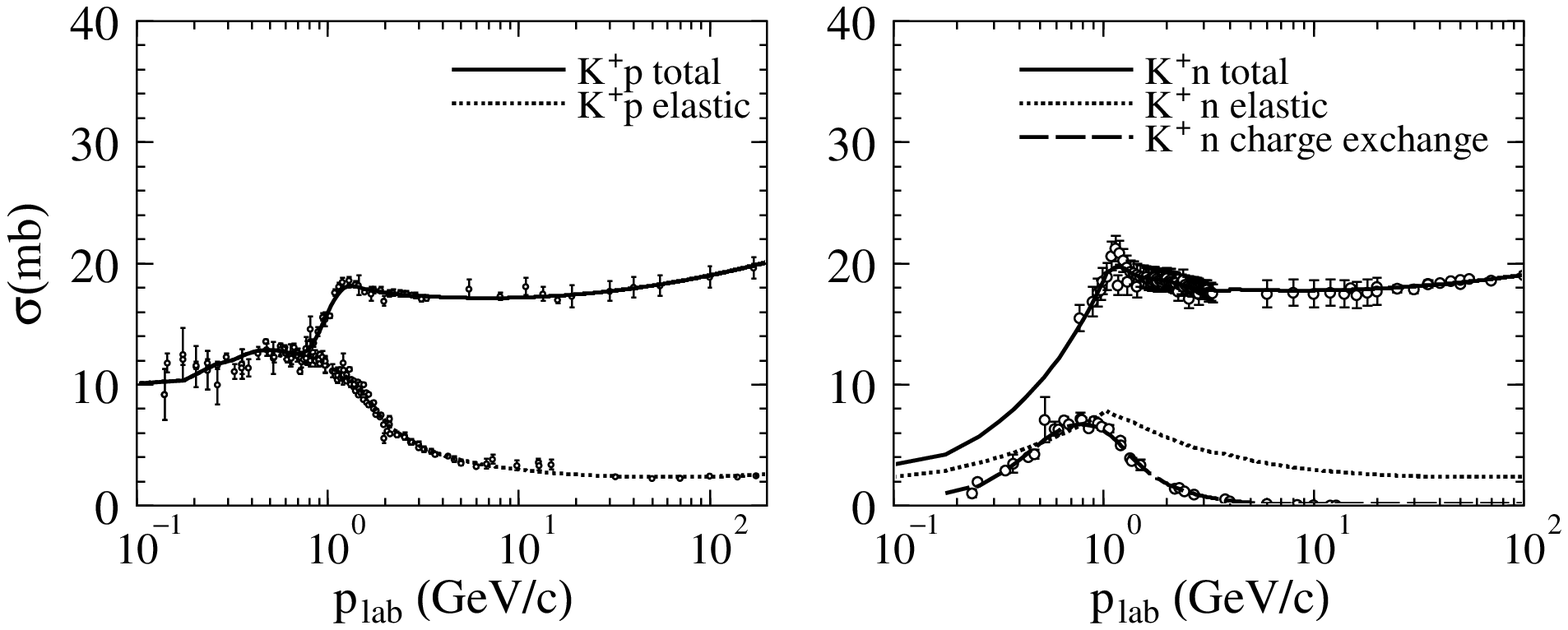,height=9cm}}}
\caption[]
      {
Parameterization of the total and elastic
$K^+ p$ and $K^+ n$ cross sections as well as charge exchange
cross section in $K^+ n$ interaction.
The data has been taken from \protect \cite{PDG96}.
     }
   \label{fig:xkaon}
\end{figure}
}

\def\FIGxkaonD{%
\begin{figure}
\centerline{\hbox{\psfig{figure=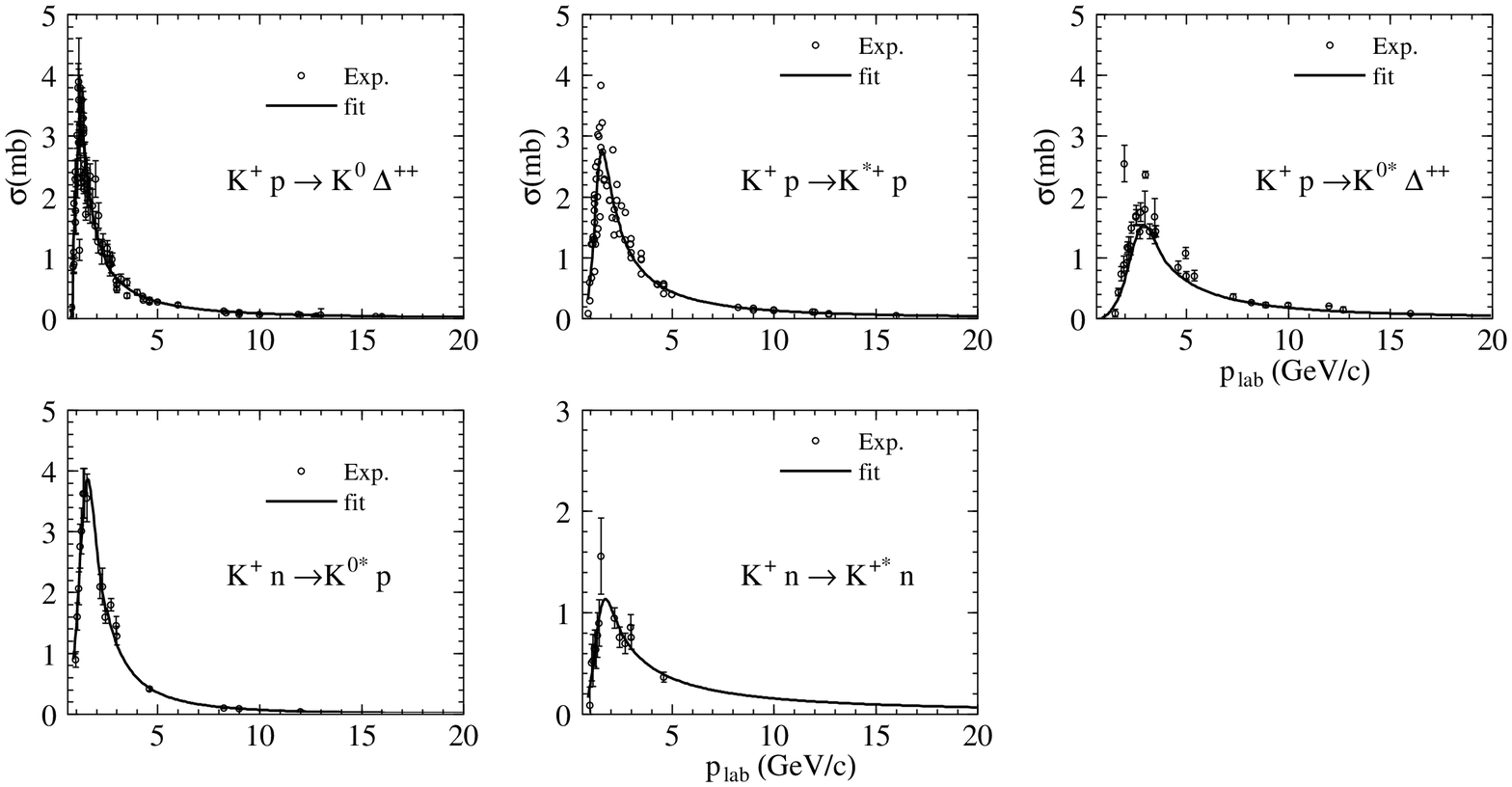,height=9.5cm}}}
\caption[]
      {
Parameterization of the $\Delta$, $K^*$ productions cross sections
in $K^+ p$ and $K^+ n$ interactions.
The data from Ref. \protect \cite{CernHera}.
     }
   \label{fig:xkaonD}
\end{figure}
}

\def\FIGppxa{%
\begin{figure}
\centerline{\hbox{\psfig{figure=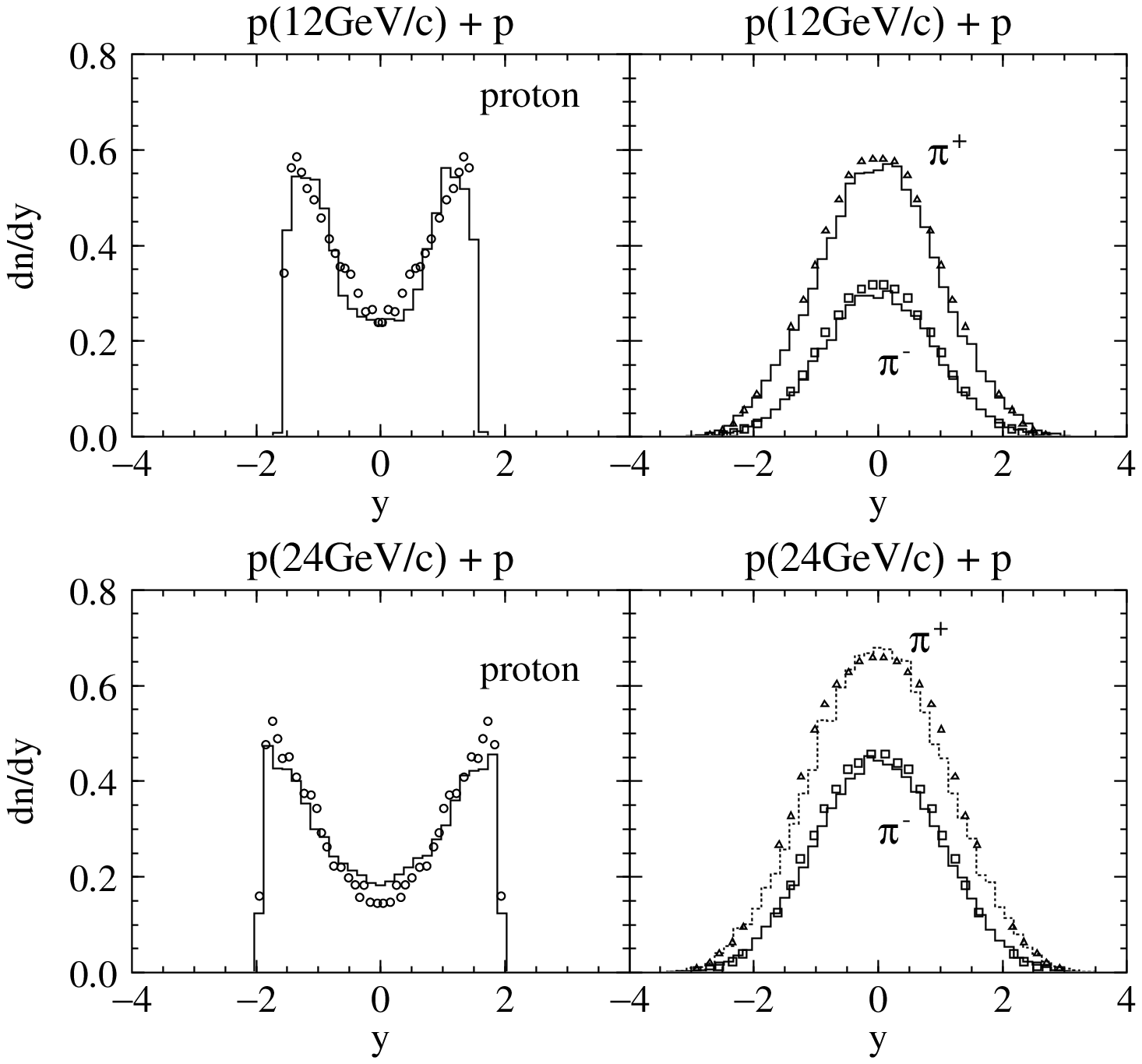,height=10cm}}}
\caption[]
      {
  The rapidity distributions
  for protons (circles), $\pi^+$ (triangles) and $\pi^-$ (squares)
   in $pp$ collisions at 12GeV/c (upper panel) and 24GeV/c (lower panel).
 Histograms are the results obtained from our calculation.
  The data are from Ref.~\cite{pp1224exp}.
     }
   \label{fig:ppxa}
\end{figure}
}

\def\FIGppxb{%
\begin{figure}
\centerline{\hbox{\psfig{figure=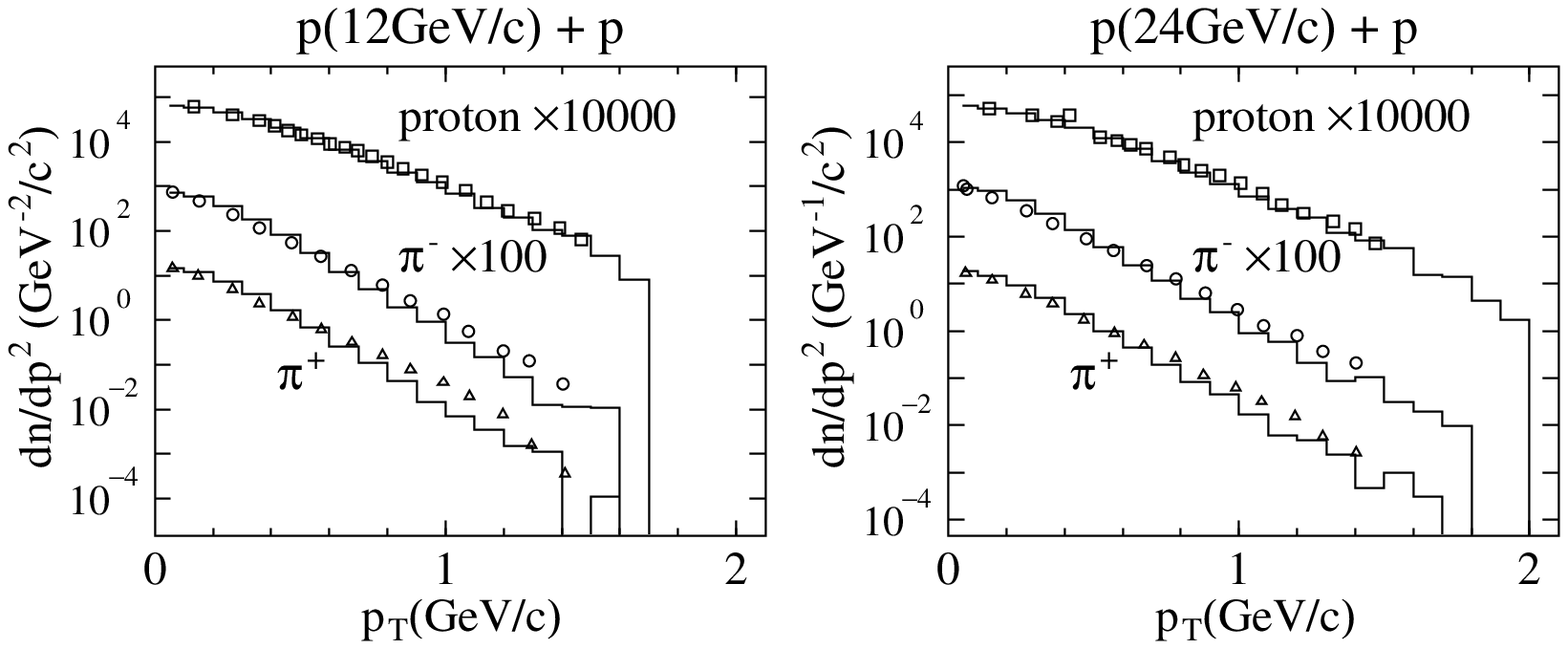,height=7cm}}}
\caption[]
      {
  The transverse momentum distributions
  for protons (squares), $\pi^+$ (triangles) and $\pi^-$ (circles)
   in $pp$ collisions at 12GeV/c(left panel) and 24GeV/c(right panel).
 Histograms are the results obtained from our calculation.
  The data are from Ref.~\cite{pp1224exp}.
     }
   \label{fig:ppxb}
\end{figure}
}

\def\FIGppx{%
\begin{figure}
\centerline{\hbox{\psfig{figure=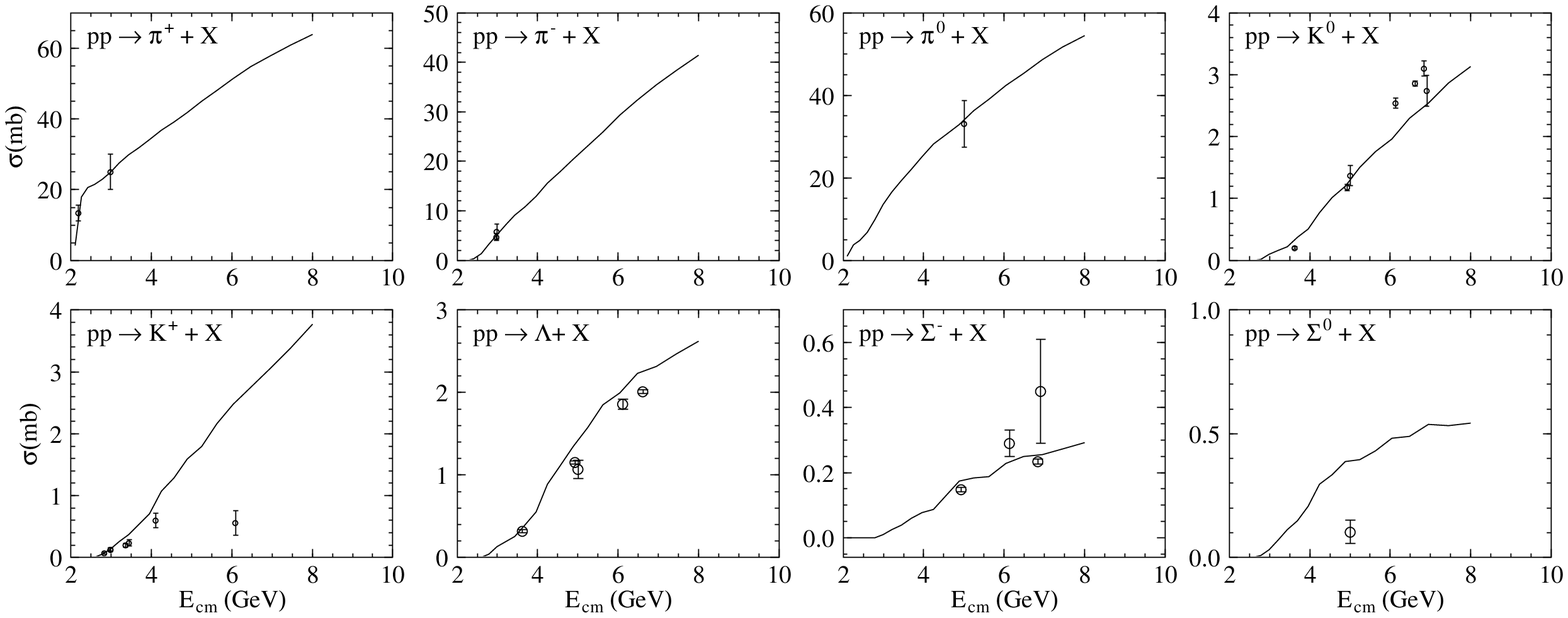,height=8cm}}}
\caption[]
      {
  The energy dependence of the inclusive production cross sections
 for pions hyperons, kaons for proton-proton
  interaction as functions of c.m. energy.
  Solid lines are the results obtained from our model.
  Data from Ref.~\cite{CernHera}.
     }
   \label{fig:ppx}
\end{figure}
}

\def\FIGproton{%
\begin{figure}
\begin{minipage}[t]{13cm}
\centerline{\hbox{\psfig{figure=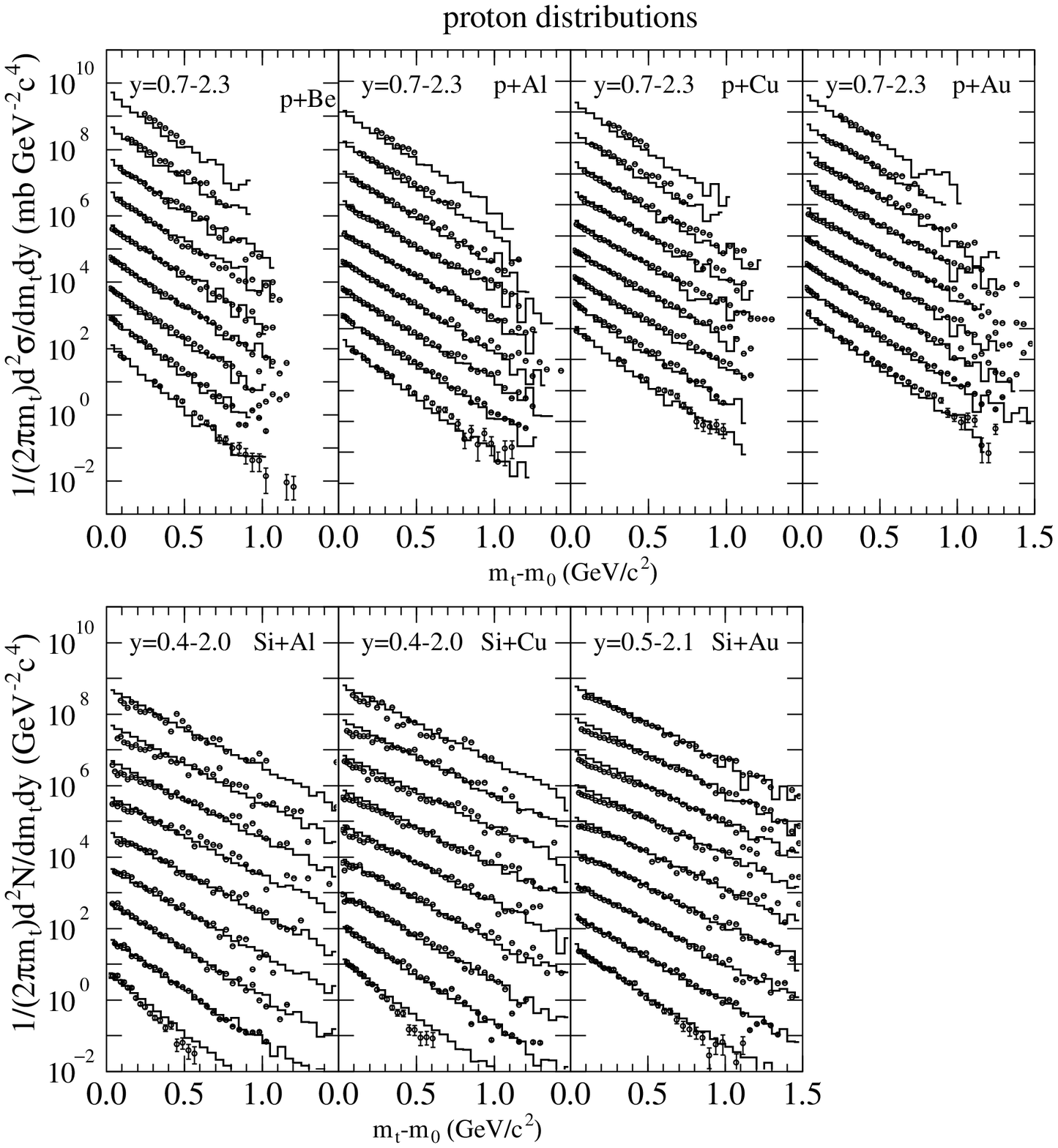,height=11cm}}}
\end{minipage}
\hfill
\begin{minipage}[t]{4.0cm}
\vspace*{-10cm}
\caption[]
      {
  Invariant cross sections of protons from
  p+Be,
  p+Al,
  p+Cu,
  p+Au
  and central Si+Al ($b\leq 1.79$fm),
  Si+Cu ($b\leq 2.2$fm),
  and Si+Au ($b\leq 2.9$fm),
  collisions  at 14.6GeV/c.
  The calculated results from Cascade model (histogram) are compared with
  the E802 data from Ref.~\cite{E802pA,E802b}.
  For proton induced collisions (upper panel), rapidity interval
  is $y=0.7$ (bottom spectrum) to $y=2.3$ (top spectrum) with
  $\delta y=0.2$.
  For Si+Al and Si+Cu collisions,
  rapidity interval is $y=0.4$ to $y=2.0$ with $\delta y=0.2$.
  For Si+Au collisions,
  rapidity interval is $y=0.4$ to $y=2.0$ with $\delta y=0.2$.
  The spectra are increased by a factor of 10 from bottom to upper.
     }
   \label{fig:proton}
\end{minipage}
\end{figure}
}
\def\FIGprotong{%
\begin{figure}
\begin{minipage}[t]{13cm}
\centerline{\hbox{\psfig{figure=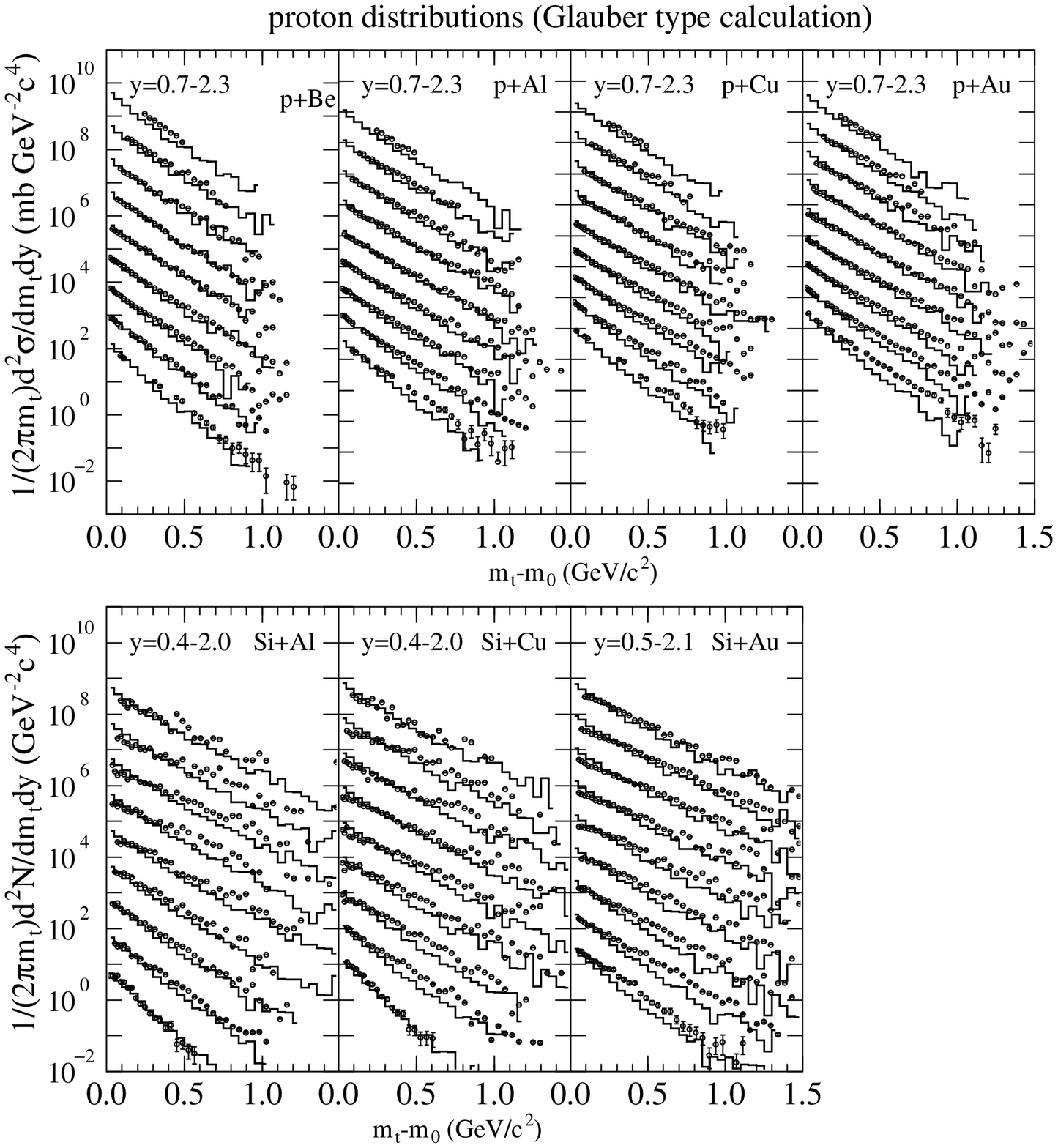,height=11cm}}}
\end{minipage}
\hfill
\begin{minipage}[t]{4.0cm}
\vspace*{-8cm}
\caption[]
      {
  The Glauber type calculations compared to the data~\cite{E802pA,E802b}.
  The meaning of the figure is the same as Fig.~\ref{fig:proton}.
     }
   \label{fig:protong}
\end{minipage}
\end{figure}
}

\def\FIGpip{%
\begin{figure}
\begin{minipage}[t]{13cm}
\centerline{\hbox{\psfig{figure=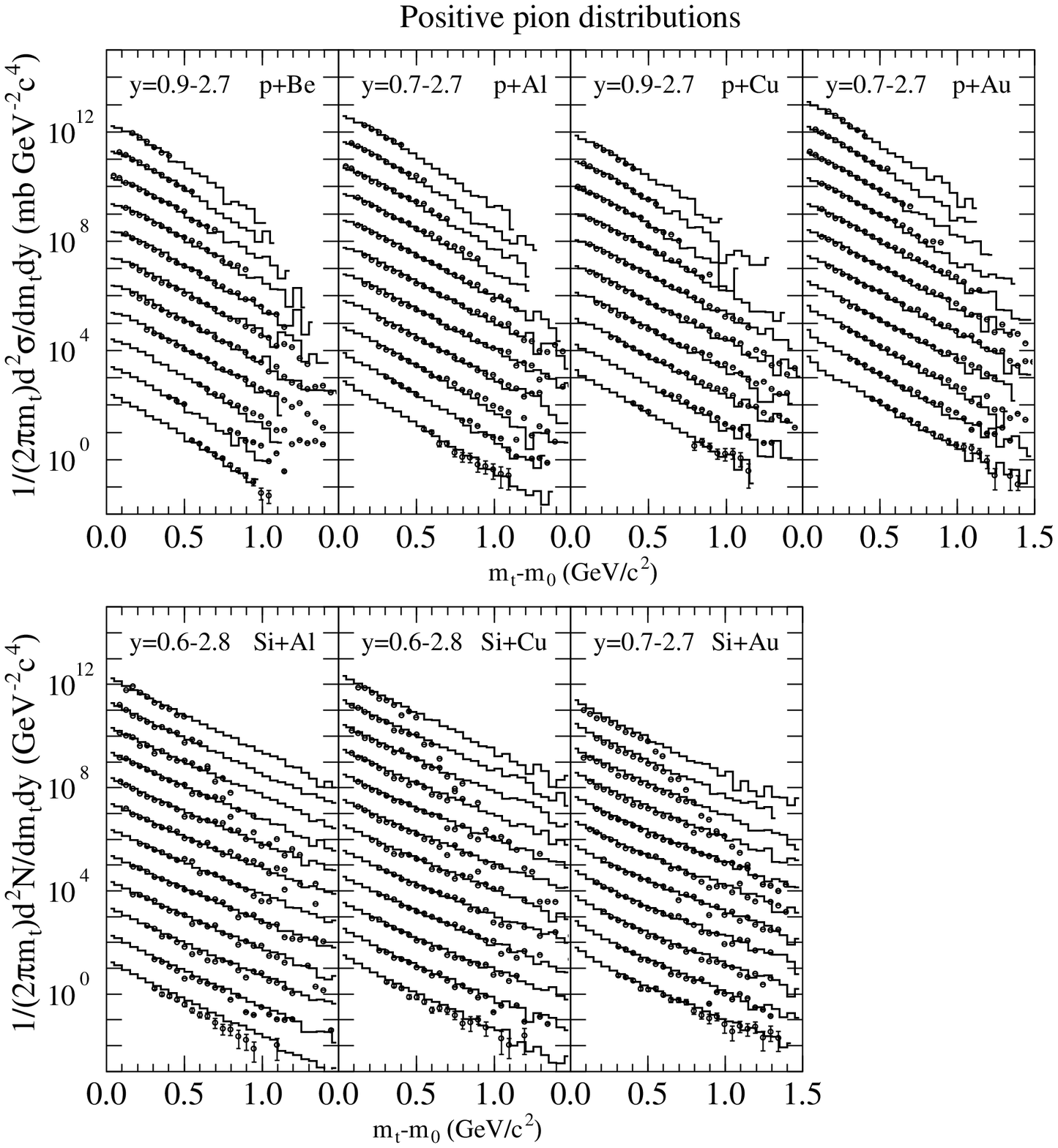,height=11cm}}}
\end{minipage}
\hfill
\begin{minipage}[t]{4.0cm}
\vspace*{-8cm}
\caption[]
      {
  Invariant cross sections of positive pions
  from p+Be ($0.9\leq y\leq 2.7$),
  p+Al ($0.7\leq y \leq 2.7$),
  p+Cu ($0.9\leq y \leq 2.7$), p+Au ($0.7\leq y \leq 2.7$),
  Si+Al ($0.6\leq y \leq 2.8$), Si+Cu ($0.6\leq y \leq 2.8$)
  and Si+Au ($0.7\leq y \leq 2.7$) collisions at 14.6GeV/c.
  The calculated results from cascade model (histogram) are compared with
  the E802 data from Ref.~\cite{E802pA,E802b}.
  The bin of rapidity is 0.2 for all collision systems.
  The spectra are increased by a factor of 10 from bottom to upper.
     }
   \label{fig:pip}
\end{minipage}
\end{figure}
}
\def\FIGpipg{%
\begin{figure}
\begin{minipage}[t]{13cm}
\centerline{\hbox{\psfig{figure=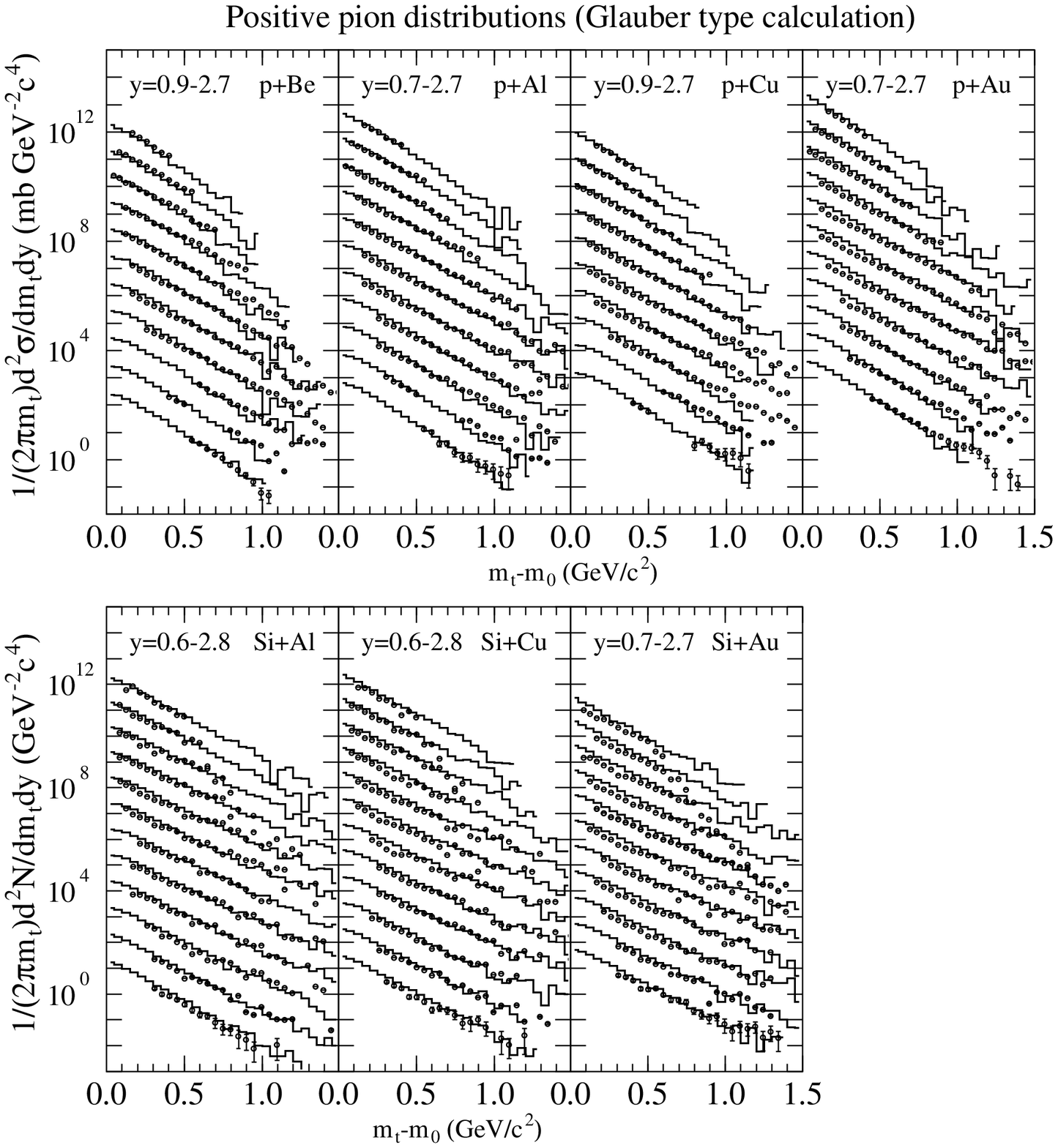,height=11cm}}}
\end{minipage}
\hfill
\begin{minipage}[t]{4.0cm}
\vspace*{-8cm}
\caption[]
      {
  Same as Fig.~\ref{fig:pip}, but the histograms show the results
  obtained from the Glauber type calculations without any rescattering
  among produced particles.
     }
   \label{fig:pipg}
\end{minipage}
\end{figure}
}
\def\FIGpin{%
\begin{figure}
\begin{minipage}[t]{13cm}
\centerline{\hbox{\psfig{figure=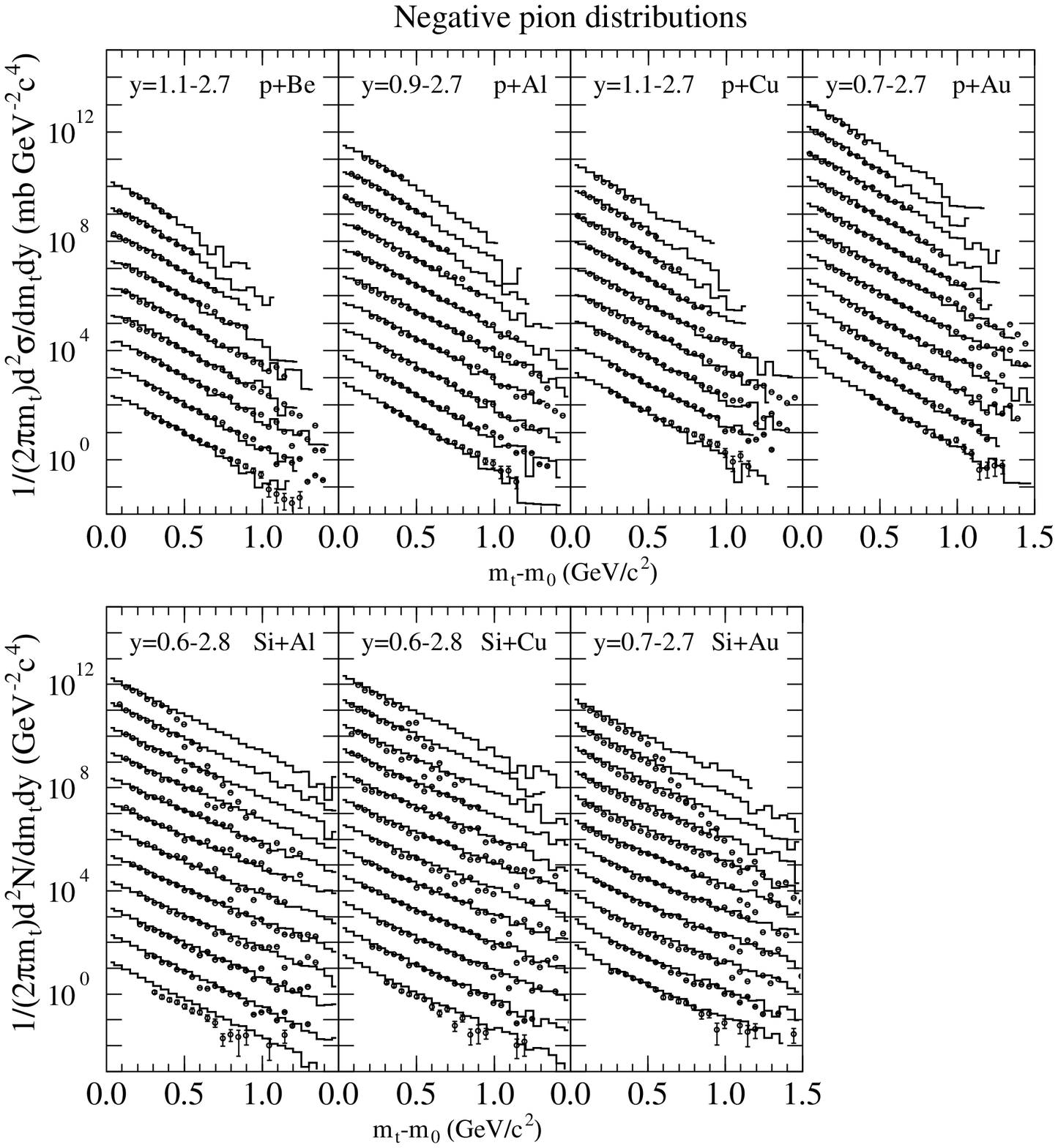,height=11cm}}}
\end{minipage}
\hfill
\begin{minipage}[t]{4.0cm}
\vspace*{-8cm}
\caption[]
      {
  Invariant cross sections of negative pions
  from p+Be ($1.1\leq y\leq 2.7$),
  p+Al ($0.9\leq y \leq 2.7$),
  p+Cu ($1.1\leq y \leq 2.7$), p+Au ($0.7\leq y \leq 2.7$),
  Si+Al ($0.6\leq y \leq 2.8$), Si+Cu ($0.6\leq y \leq 2.8$)
  and Si+Au ($0.7\leq y \leq 2.7$) collisions at 14.6GeV/c.
  The calculated results from Cascade model (histogram) are compared with
  the E802 data from Ref.~\cite{E802pA,E802b}.
  The bin of rapidity is 0.2 for all collision systems.
  The spectra are increased by a factor of 10 from bottom to upper.
     }
   \label{fig:pin}
\end{minipage}
\end{figure}
}
\def\FIGping{%
\begin{figure}
\begin{minipage}[t]{13cm}
\centerline{\hbox{\psfig{figure=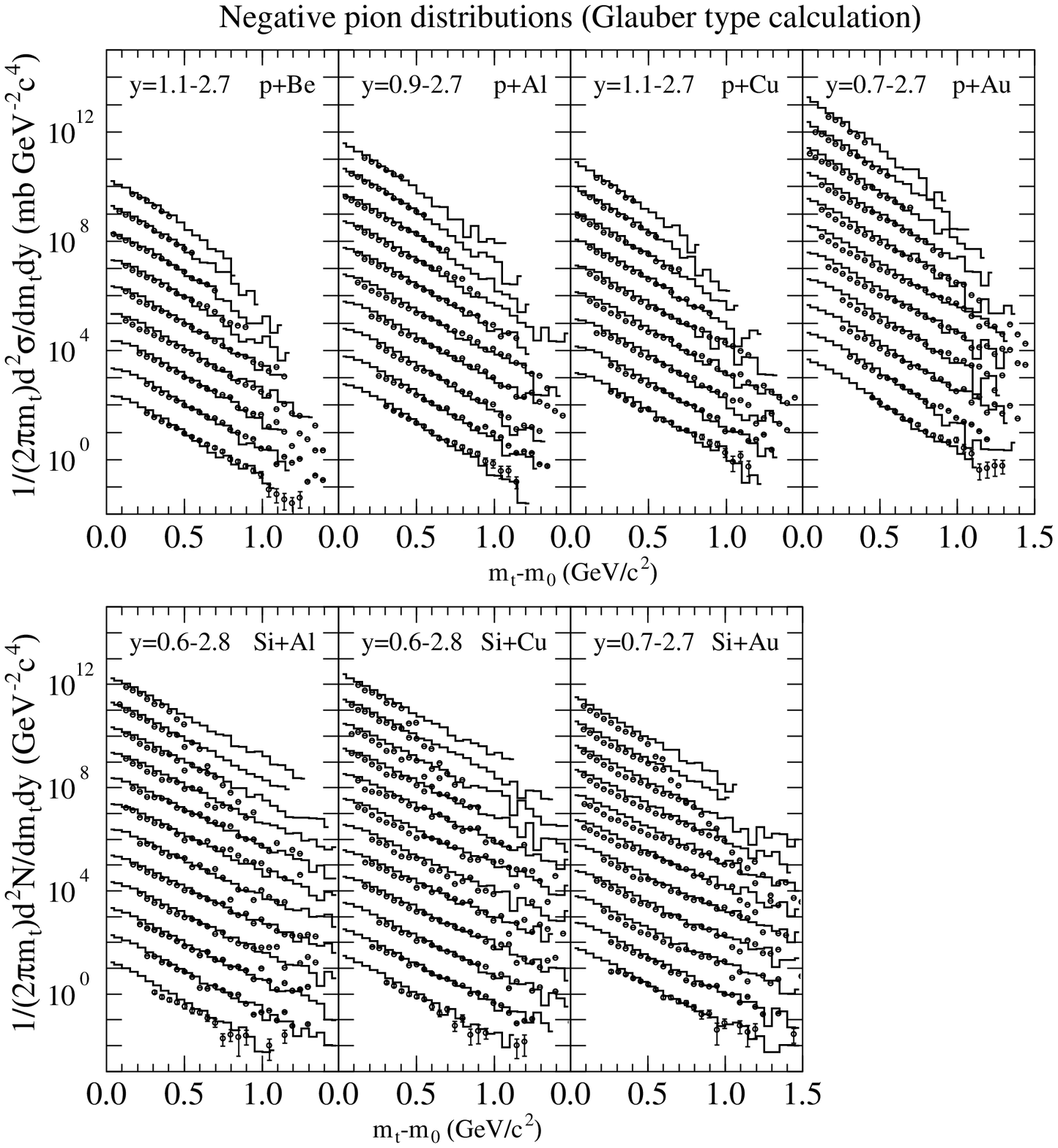,height=11cm}}}
\end{minipage}
\hfill
\begin{minipage}[t]{4.0cm}
\vspace*{-8cm}
\caption[]
      {
  Same as Fig.~\ref{fig:pin}, but the histograms show the results
  obtained from the Glauber type calculations without any rescattering
  among produced particles.
     }
   \label{fig:ping}
\end{minipage}
\end{figure}
}

\def\FIGkp{%
\begin{figure}
\begin{minipage}[t]{13cm}
\centerline{\hbox{\psfig{figure=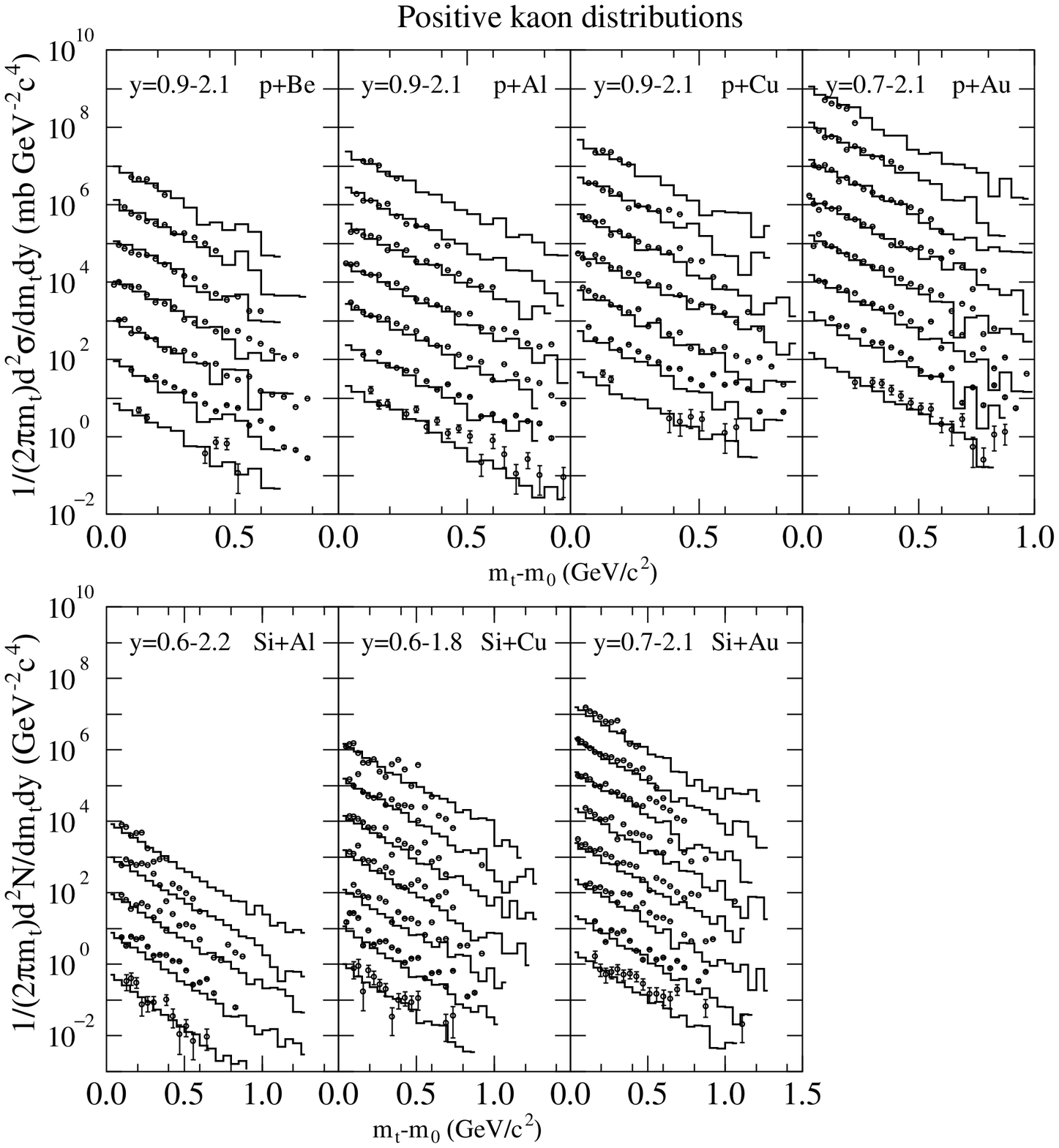,height=11cm}}}
\end{minipage}
\hfill
\begin{minipage}[t]{4.0cm}
\vspace*{-8cm}
\caption[]
      {
  Comparison of the invariant transverse momentum spectra of
  $K^+$
  between the cascade model and the experimental data of ~\cite{E802pA,E802b}
 in p+Be ($0.9\leq y \leq 2.1$ with bin size 0.2),
   p+Al ($0.9\leq y \leq 2.1$ with bin size 0.2),
   p+Cu ($0.9\leq y \leq 2.1$ with bin size 0.2),
   p+Au ($0.7\leq y \leq 2.1$ with bin size 0.2),
  Si+Al ($0.6\leq y \leq 2.2$ with bin size 0.2),
  Si+Cu ($0.6\leq y \leq 1.8$ with bin size 0.4) and
  Si+Au ($0.7\leq y \leq 2.1$ with bin size 0.2) collisions at 14.6GeV/c.
  The spectra are increased by a factor of 10 from bottom to upper.
     }
   \label{fig:kp}
\end{minipage}
\end{figure}
}

\def\FIGkn{%
\begin{figure}
\begin{minipage}[t]{13cm}
\centerline{\hbox{\psfig{figure=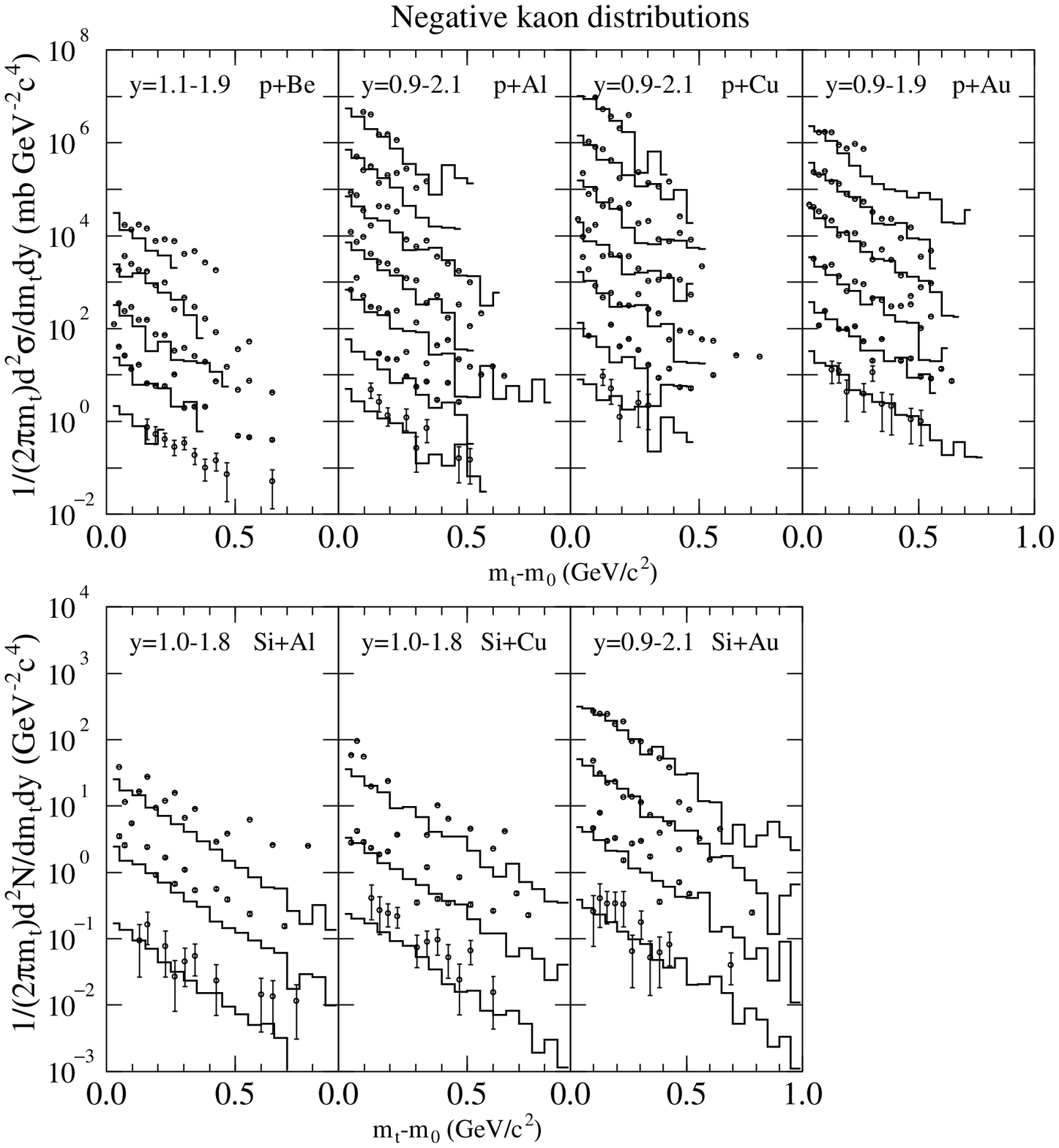,height=11cm}}}
\end{minipage}
\hfill
\begin{minipage}[t]{4.0cm}
\vspace*{-8cm}
\caption[]
      {
  Comparison of the invariant transverse momentum spectra of
 $K^-$ 
 in p+Be ($1.1\leq y \leq 1.9$ with bin size 0.2),
   p+Al ($0.9\leq y \leq 2.1$ with bin size 0.2),
   p+Cu ($0.9\leq y \leq 2.1$ with bin size 0.2),
   p+Au ($0.9\leq y \leq 1.9$ with bin size 0.2),
  Si+Al ($1.0\leq y \leq 1.8$ with bin size 0.4),
  Si+Cu ($1.0\leq y \leq 1.8$ with bin size 0.4) and
  Si+Au ($0.9\leq y \leq 2.1$ with bin size 0.4) collisions at 14.6GeV/c.
  The spectra are increased by a factor of 10 from bottom to upper.
     }
   \label{fig:kn}
\end{minipage}
\end{figure}
}

\def\FIGkpg{%
\begin{figure}
\begin{minipage}[t]{13cm}
\centerline{\hbox{\psfig{figure=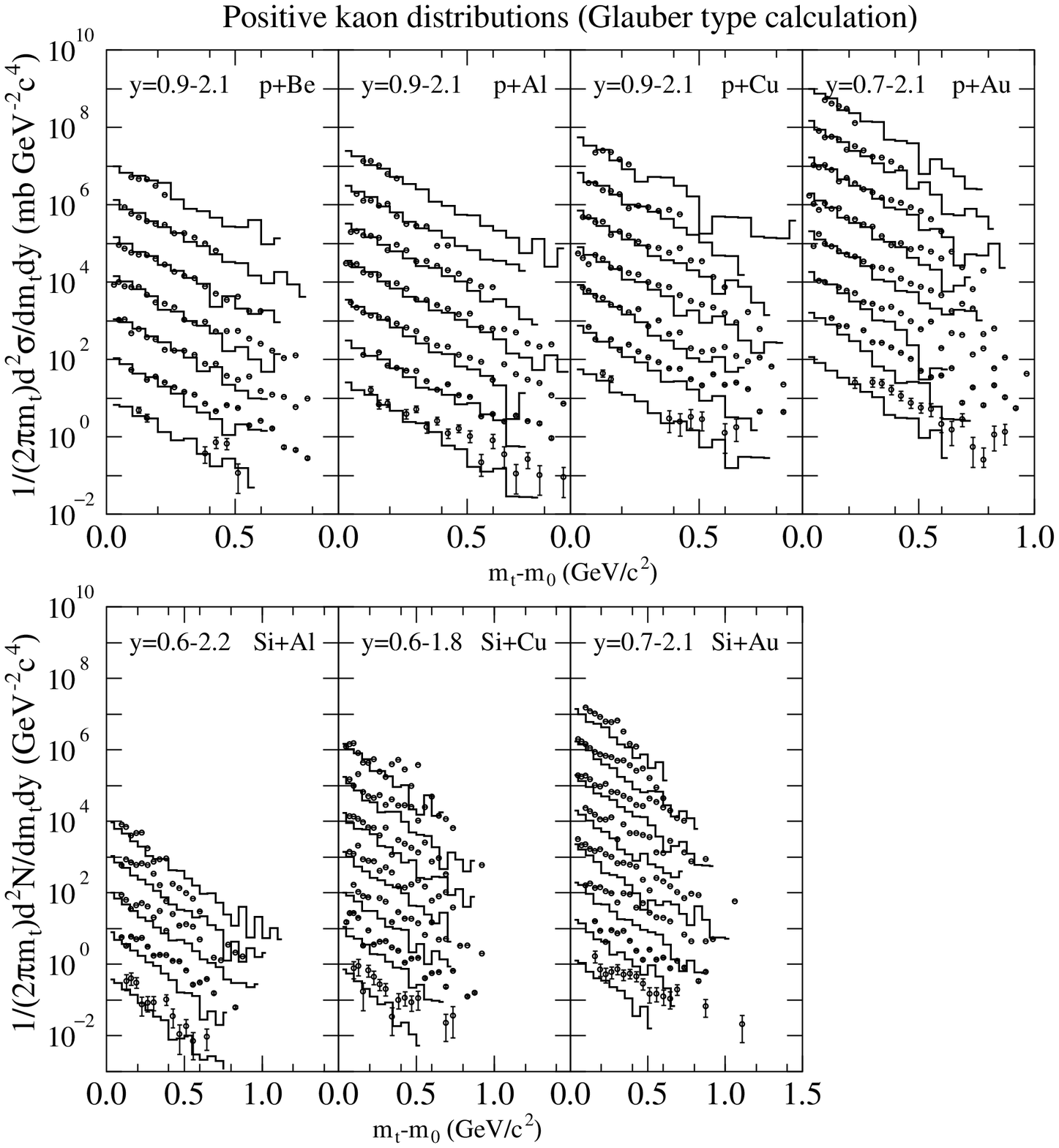,height=11cm}}}
\end{minipage}
\hfill
\begin{minipage}[t]{4.0cm}
\vspace*{-8cm}
\caption[]
      {
  Glauber type calculations of
  invariant cross sections of $K^+$
  for p+Be, p+Al, p+Cu, p+Au, Si+Al, Si+Cu and Si+Au reactions at 14.6GeV/c.
  in comparison to the E802 data from~\cite{E802pA,E802b}.
  The meaning of figure is the same as Fig.~\ref{fig:kp}.
     }
   \label{fig:kpg}
\end{minipage}
\end{figure}
}

\def\FIGkng{%
\begin{figure}
\begin{minipage}[t]{13cm}
\centerline{\hbox{\psfig{figure=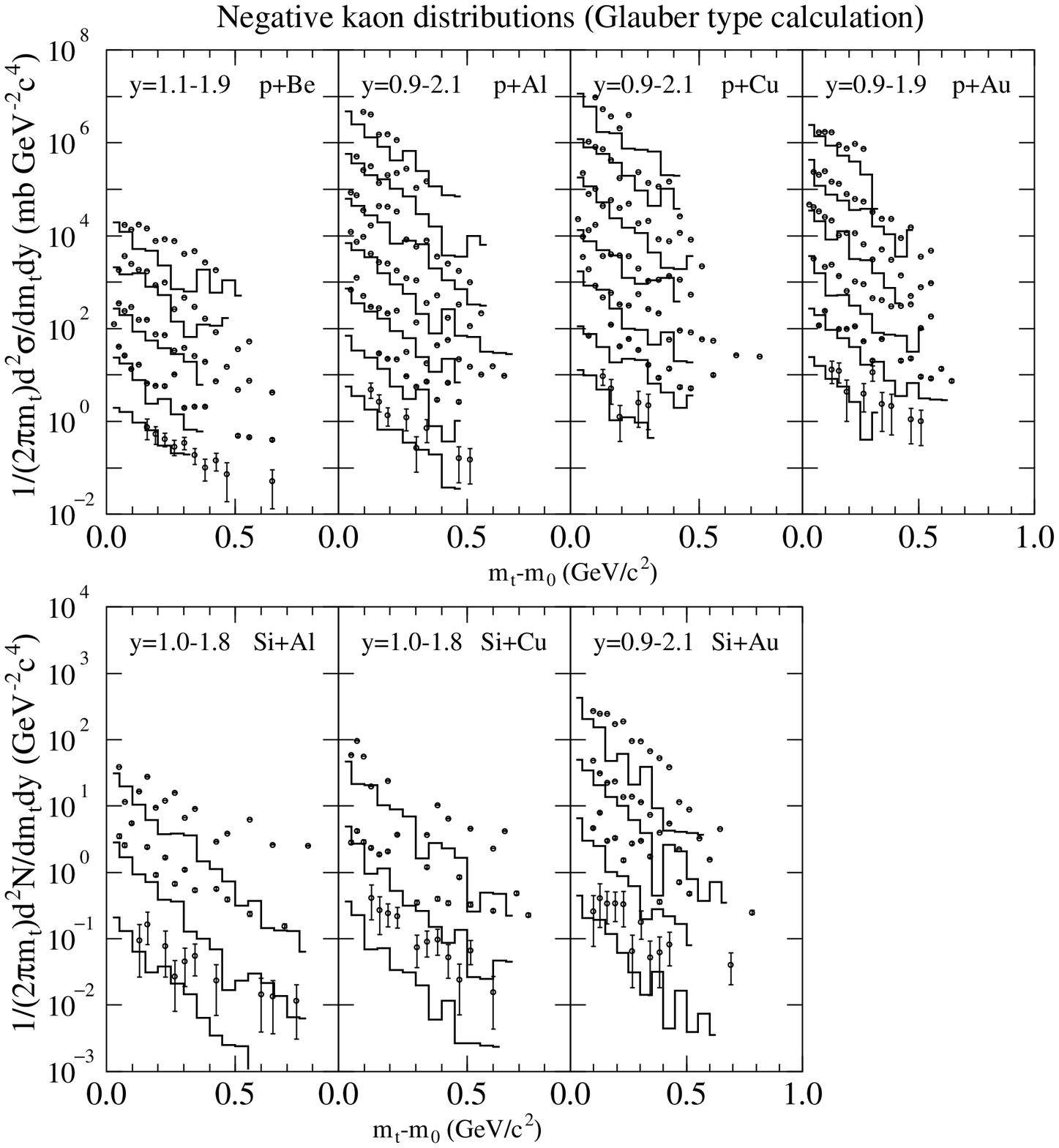,height=11cm}}}
\end{minipage}
\hfill
\begin{minipage}[t]{4.0cm}
\vspace*{-8cm}
\caption[]
      {
  Glauber type calculations of
  invariant cross sections of $K^-$
  for p+Be, p+Al, p+Cu, p+Au, Si+Al, Si+Cu and Si+Au reactions at 14.6GeV/c.
  in comparison to the E802 data from~\cite{E802pA,E802b}.
  The meaning of figure is the same as Fig.~\ref{fig:kn}.
     }
   \label{fig:kng}
\end{minipage}
\end{figure}
}

\def\FIGAuAumtp{%
\begin{figure}
\begin{minipage}[t]{13cm}
\centerline{\hbox{\psfig{figure=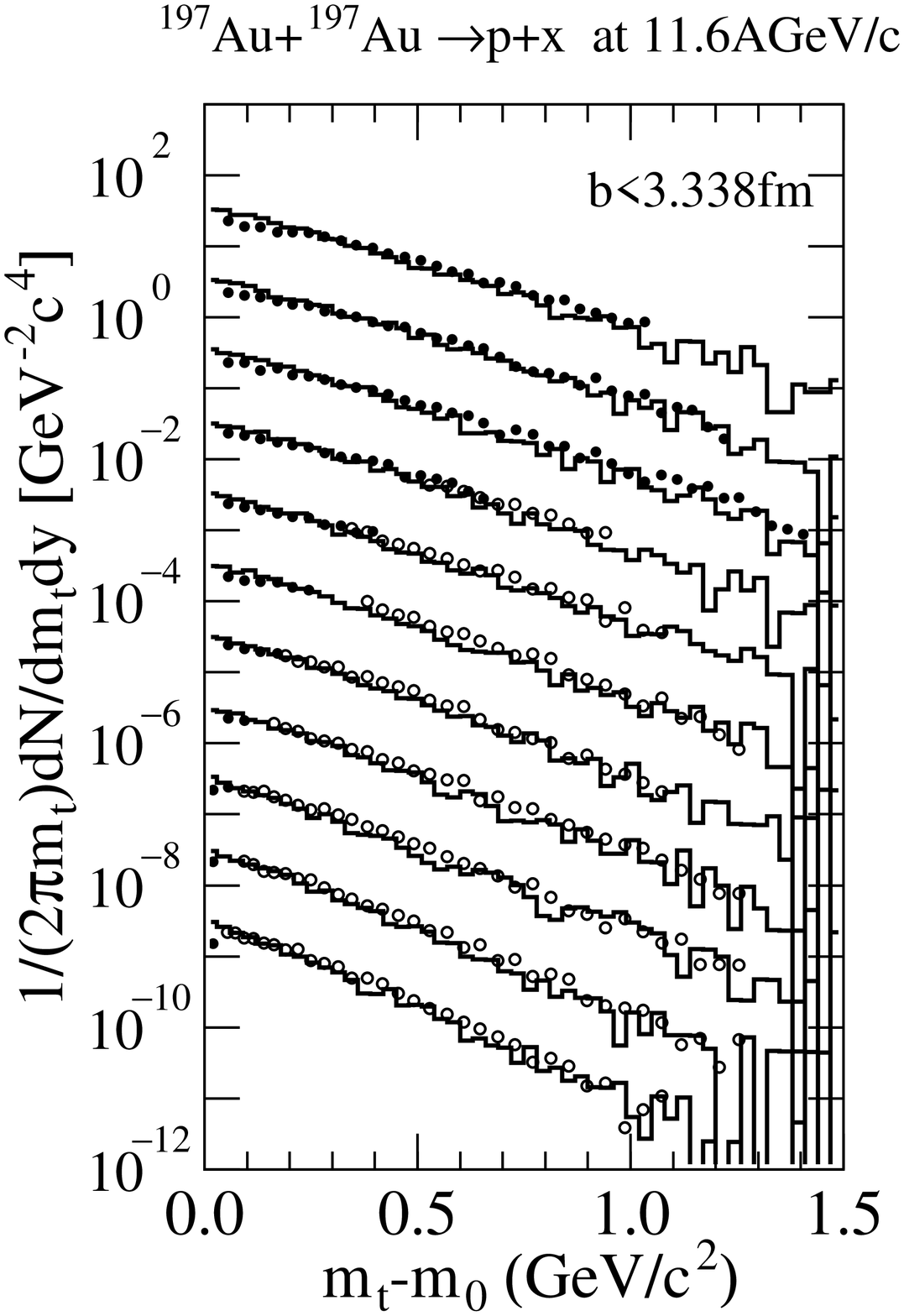,height=8cm}}}
\end{minipage}
\hfill
\begin{minipage}[t]{4.0cm}
\vspace*{-7cm}
\caption[]
      {
  The result of Cascade model calculation of
 transverse mass distributions of protons for
  central Au+Au collision at 11.6GeV/c
 in different rapidity intervals. The spectra are scaled down by a
 factor of 10 successively from upper corresponds to the 
 c.m. rapidity $y=0.05$ to lower spectrum of $y=1.05$ with
 the bin width of 0.1. 
 Impact parameter is distributed from 0 to 3.338fm.
 Data are from Ref.~\cite{E802AuAu}.
     }
   \label{fig:AuAumtp}
\end{minipage}
\end{figure}
}

\def\FIGAuAurapmt{%
\begin{figure}
\begin{minipage}[t]{8.5cm}
\centerline{\hbox{\psfig{figure=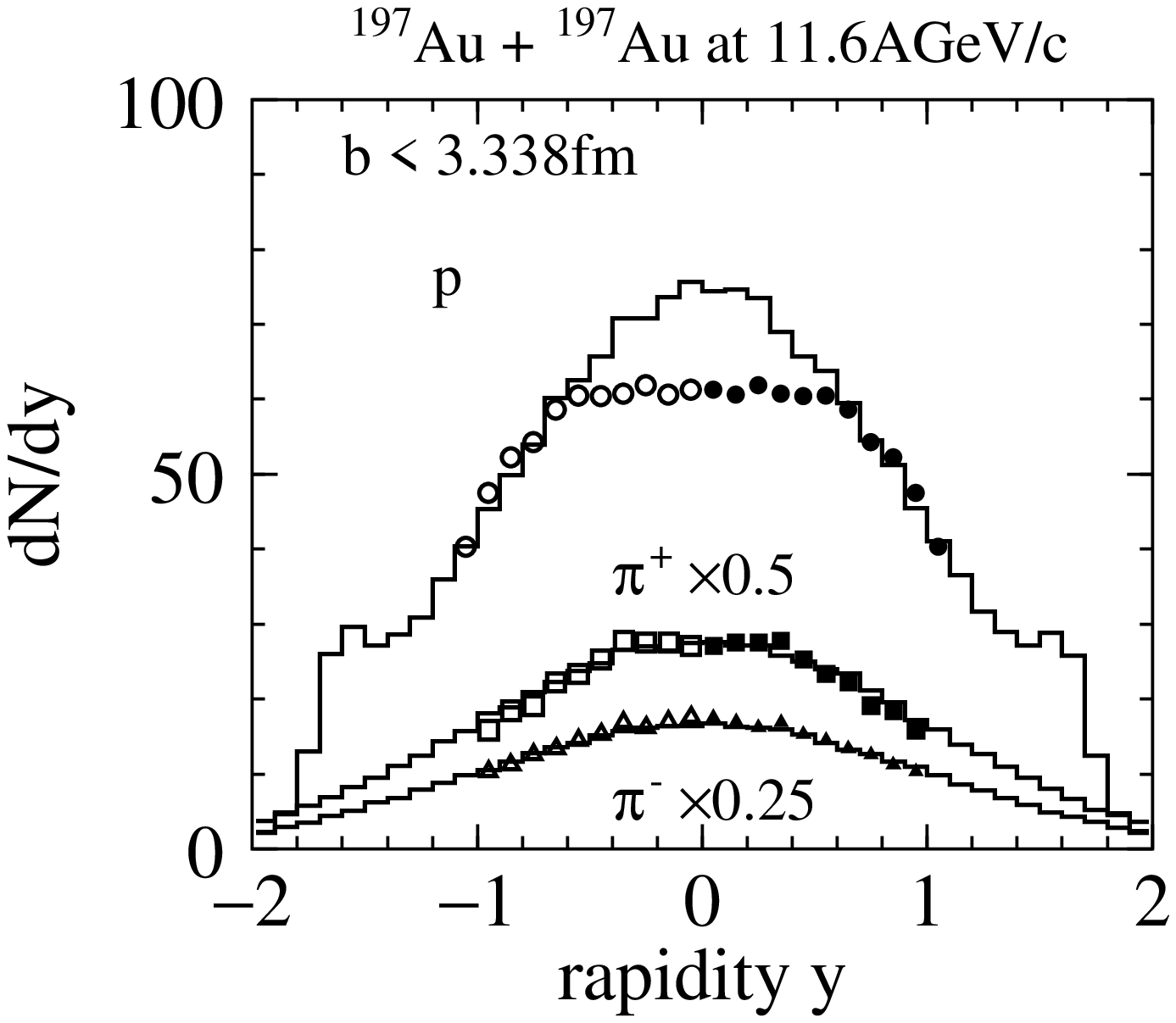,height=8cm}}}
\caption[]
      {
  Comparison of 
  rapidity distributions of protons (circles), positive pions (squares)
  and negative pions (triangles)
  between experimental data~\cite{E802AuAu} and cascade
  model calculations (histograms) for central Au+Au collision at 11.6GeV/c.
  The data for positive pions  is scaled down by a factor of 0.5
  and for negative pions 0.25.
  In cascade calculations, impact parameter is distributed from 0 to 3.338fm.
     }
   \label{fig:AuAurap}
\end{minipage}
\hfill
\begin{minipage}[t]{8.5cm}
\centerline{\hbox{\psfig{figure=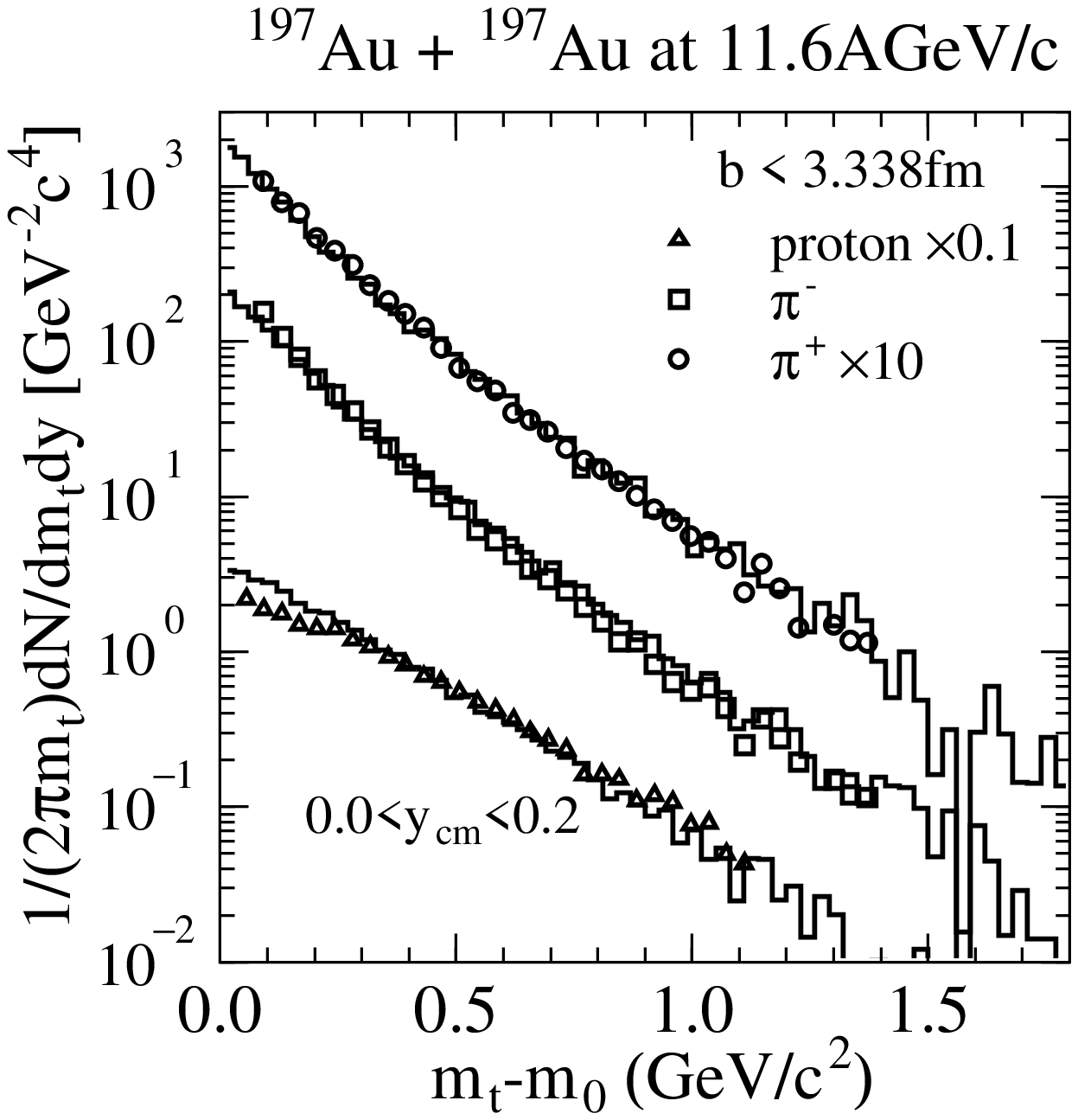,height=8cm}}}
\caption[]
      {
  Transverse mass distributions of protons and pions for
  central Au+Au collision at 11.6GeV/c.
  The triangles represent the data of protons  scaled by a factor of 0.1
  squares correspond to $\pi^-$ data
 and circles $\pi^+$ data scaled by a factor of 10
 from from Ref.~\cite{E802AuAu}.
 Histograms represent the results from Cascade model
 with impact parameter $b\leq 3.338$fm.
     }
   \label{fig:AuAumt}
\end{minipage}
\end{figure}
}

\def\FIGncoll{%
\begin{figure}
\centerline{\hbox{\psfig{figure=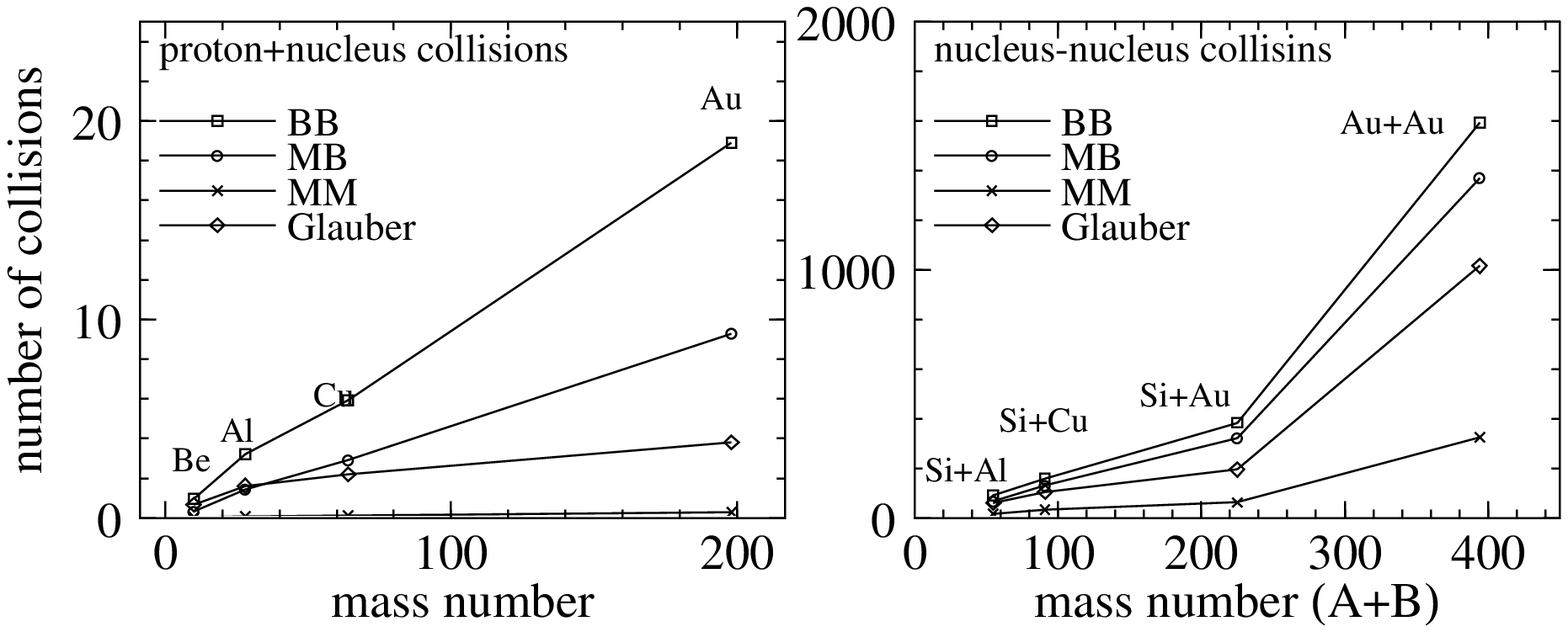,height=8cm}}}
\caption[]
      {
  Mass dependence of number of $BB$ (open squares),
  $MB$ (open circles) and $MM$ (open crosses) collisions
   obtained from cascade model calculations and Glauber type calculations
  (open diamonds) for p+A collisions in the left panel
  and for Si+A and Au+Au collisions (right panel).
     }
   \label{fig:ncoll}
\end{figure}
}

\def\FIGmesons{%
\begin{figure}
\centerline{\hbox{\psfig{figure=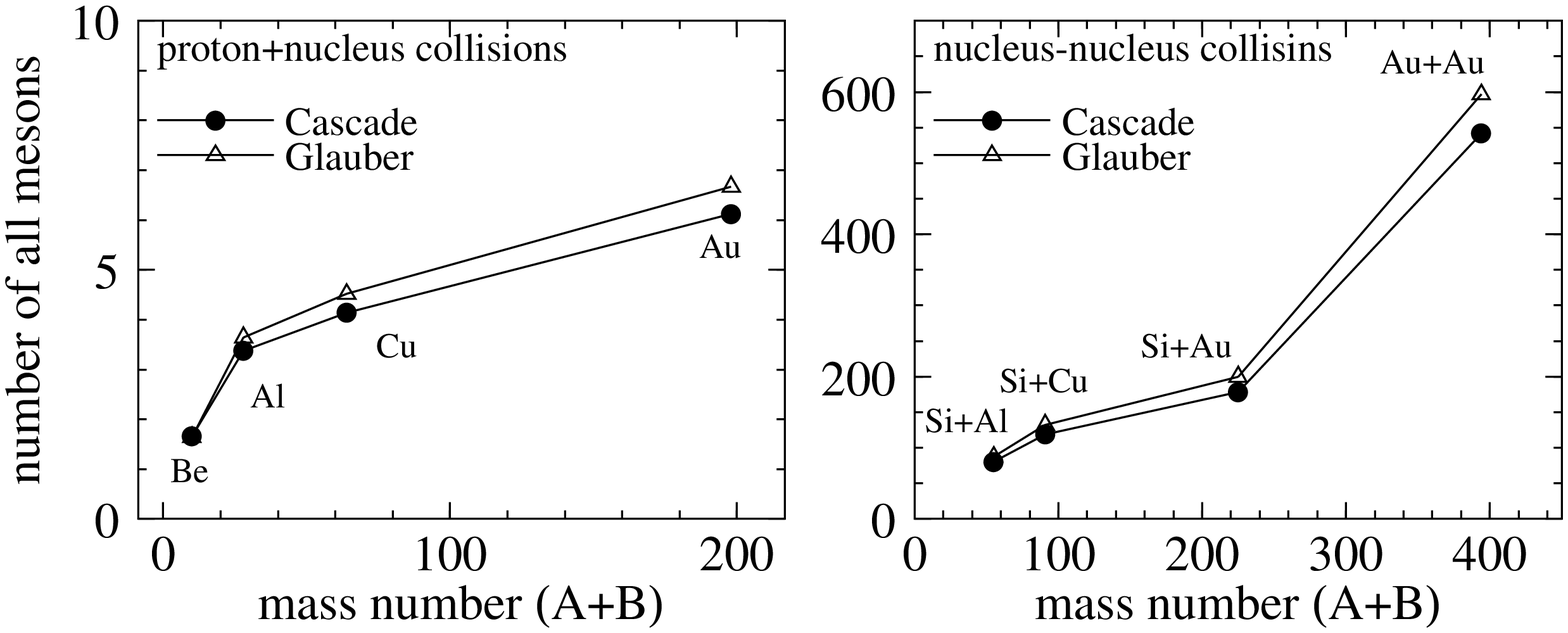,height=8cm}}}
\caption[]
      {
  Mass dependence of number of total mesons
  from cascade model calculations (full circles)
  and Glauber type calculations (full triangles)
  for Si+A (A=Al, Cu, Au), Au+Au collisions (right panel)
  and for p+A (A=Be, Al, Cu, Au) collisions (left panel).
  Meson multiplicity is reduced by including rescattering.
     }
   \label{fig:mesons}
\end{figure}
}

\def\FIGcolls{%
\begin{figure}
\centerline{\hbox{\psfig{figure=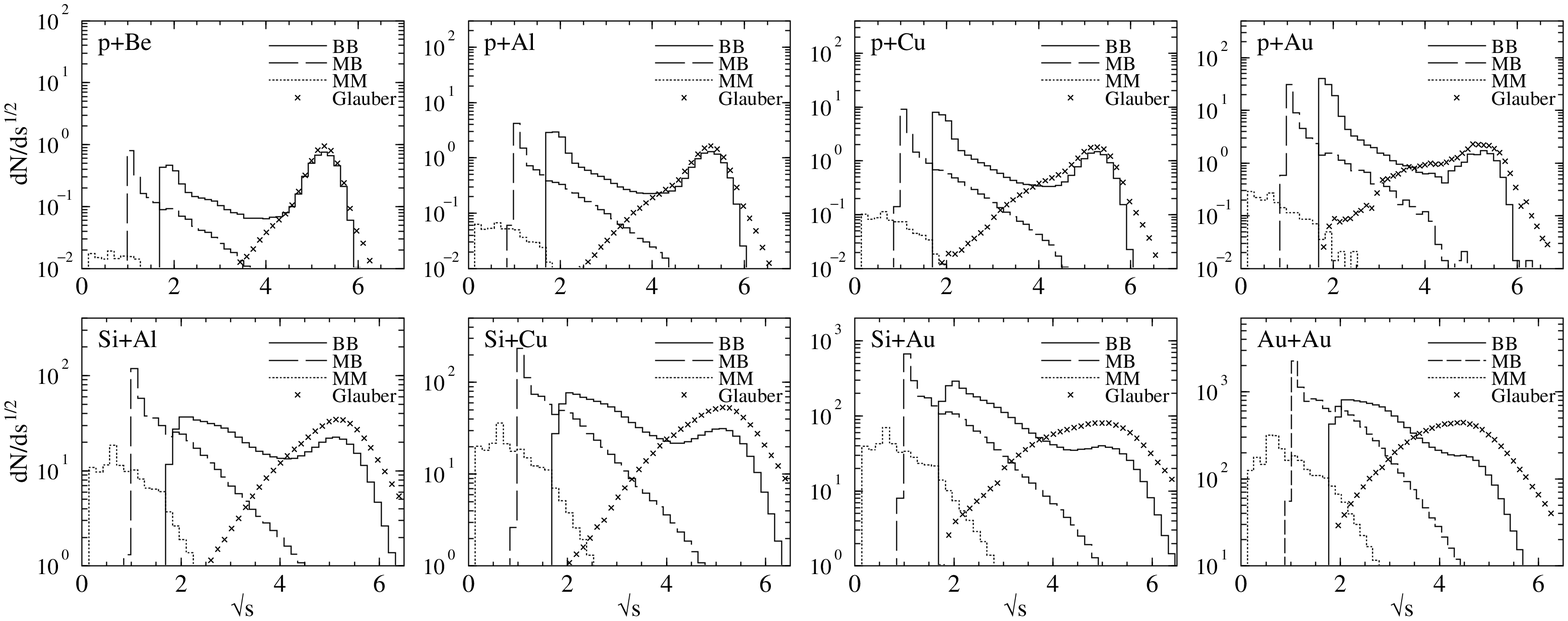,height=8cm}}}
\caption[]
      {
  Collision spectrum of $BB$ (full lines), $MB$ (dashed lines)
  and $MM$ (dotted lines) collisions
  from the cascade model calculations and Glauber type calculations
  (crosses)
  for the p+Be, p+Al, p+Cu, p+Au, Si+Al, Si+Cu, Si+Au and Au+Au
  collisions at AGS energies.
     }
   \label{fig:colls}
\end{figure}
}

\psfigurepath{Fig}

\begin{document}
\title{Study of relativistic nuclear collisions at AGS energies
        from p+Be to Au+Au with hadronic cascade model}

\author{Y. Nara$^{1,2}$, N. Otuka$^3$, A. Ohnishi$^3$,
         K. Niita$^1$ and S. Chiba$^1$}

\address{$^1$ Advanced Science Research Center, 
        Japan Atomic Energy Research Institute,
        Tokai, Naka, Ibaraki 319-1195, Japan}
\address{$^2$ Physics Department, Brookhaven National Laboratory,
         Upton, N.Y. 11973, U.S.A.}
\address{$^3$    Division of Physics, Graduate School of Science,
        Hokkaido University, Sapporo 060-0810, Japan}

\maketitle

\begin{abstract}
 A hadronic cascade model
 based on resonances and strings
  is used to study mass dependence of relativistic nuclear collisions from
  p+Be to Au+Au at AGS energies ($\sim 10\AGeV$) systematically.
Hadron transverse momentum and rapidity distributions
 obtained with both cascade calculations and Glauber type calculations
are compared with experimental data
 to perform detailed discussion about the importance of rescattering
 among hadrons.
We find good agreement with the experimental data
without any change of model parameters with the cascade model.
It is found that rescattering is of importance both
 for the explanation of high transverse momentum tail
 and for the multiplicity of produced particles.
\end{abstract}


\section{INTRODUCTION}

There is now a big interest in
 studying strongly interacting matter
 at high density and/or temperature
   which is created in high energy nuclear collisions.
Indeed,
at high densities and/or temperatures, QCD predicts 
 the chiral symmetry restoration and quark deconfinement.
How can we create such matter ?
At present,
  high energy heavy ion collision is considered to be a unique way
 to create such dense and hot matter at laboratory.
In order to find such a new form of nuclear matter, 
several heavy ion experiments have been and are being performed
 with Si(14.6\AGeVc) or Au(11.6\AGeVc) beam at BNL-AGS
 and with the O(200\AGeVc), S(200\AGeVc) or Pb(158\AGeVc) beam at CERN-SPS.

Since high energy heavy ion collisions
 lead to a huge number of final states,
 many event generators have been proposed
 to explore these high energy nuclear collisions,
 with the aid of Monte-Carlo realization of complex processes.
In these event generators,
 there are mainly three categories of models.
%
%
%
The models in the first category
 assume Glauber geometry for the treatment of AA collisions.
For example,
FRITIOF~\cite{fritiof}, LUCIAE~\cite{luciae}, VENUS~\cite{venus},
HIJING~\cite{hijing}, DPM~\cite{dpm}, HIJET~\cite{hijet} and LEXUS\cite{lexus}
belong to this category.
Final interaction among hadrons are included in VENUS, HIJET and LUCIAE.
In these models, main quantum features during the multiple scattering
are preserved within the eikonal approximation,
and efficiently fast calculations are possible.
However, these approaches are mainly designed for the extremely high
energy collisions ($\srt>10\AGeV$).

%
%
%
The models in the second category (parton cascade models), 
 such as VNI~\cite{vni} and ZPC~\cite{zpc}, have been recently developed
to implement the interaction among partons to study space-time
evolution of partons produced in high energy nuclear collisions.
%
These models have been originally designed
 to describe ultra-relativistic heavy ion collisions at collider energies,
 such as BNL-RHIC and CERN-LHC,
and they have met some successes in describing heavy ion collisions
  at CERN-SPS energies~\cite{kkg}.
%

%
%
%
The third category of models is a
transport model which is often referred to as 'hadronic cascade'.
For example, RQMD~\cite{rqmd1,rqmd2}, QGSM~\cite{qgsm}, ARC~\cite{arc},
  ART~\cite{art}, UrQMD~\cite{urqmd} and HSD~\cite{hsd}
  can be categorized here.
They have been successfully used to describe many aspects of
 high energy heavy ion collisions in a wide range of incident energies.
For the description of AA collisions in hadronic cascade models,
 the trajectories of all hadrons as well as resonances
 including produced particles are followed explicitly
   as a function of time.
Nuclear collisions are modeled by the sum of independent
hadron-hadron ($hh$) collisions without interferences.
Two particles are made to collide if their closest distance is smaller than
$\sqrt{\sigma(s)/\pi}$, here $\sigma(s)$ represents the total cross
section at the c.m. energy $\srt$.
As a result of the $hh$ collision,
 secondary particles will be produced according to
the specific model with some formation time.
 One of the most distinct difference among these models
 may be in the way of implementing hadronic degrees of freedom.
 In RQMD and UrQMD, many established hadronic resonances
 are explicitly propagated in space-time, while ARC, ART and HSD
 do not include higher hadronic resonances.
Although both modelings seem to give similar results
 if we see the final hadronic spectra inclusively,
 we expect thermodynamic quantities like pressure or temperature
 before freeze-out predicted by those models
would be different from each other~\cite{YNOM}.
Another difference is the treatment of multiparticle production.
String model is adapted in RQMD, QGSM, UrQMD and HSD, while
in ARC and ART, final states are sampled according to the
direct parameterization of the experimental data.
The hadronic cascade model based on the string phenomenology
implies that some partonic degrees of freedom play 
 some roles in reaction dynamics implicitly.
In fact, the estimation of partonic degrees of freedom has been done
recently within UrQMD~\cite{urqmd2}.
ARC~\cite{arc} has shown that
  'pure' hadronic model can describe the data at AGS energies.
At collider energies, however, explicit treatments of
 partonic degrees of freedom will be necessary.

%
%
The main purpose of this work
   is 
  to perform systematic analyses of collisions
   from pA to massive AA systems at AGS energies,
   for which high-quality systematic experimental data
   are available~\cite{E802pA,E802b},
   within the hadronic cascade model, JAM1.0,
   which has been developed recently
   based on resonances, strings and pQCD.

The main features included in JAM are as follows.
 (1) At low energies, inelastic $hh$ collisions are modeled by
     the resonance productions based on the idea from RQMD and UrQMD.
 (2) Above the resonance region, soft string excitation is implemented
     along the lines of the HIJING model~\cite{hijing}.
 (3) Multiple minijet production is also included in the same way
     as the HIJING model in which jet cross section and the number
     of jet are calculated using an eikonal formalism for
     perturbative QCD (pQCD) and hard parton-parton scatterings
     with initial and final state radiation are simulated
     using PYTHIA~\cite{pythia} program.
 (4) Rescattering of hadrons which have original constituent quarks can
     occur with other hadrons
     assuming the additive quark cross section within a formation time.
Since these features of the present hadronic cascade model, JAM1.0,
enables us to explore heavy ion collisions in a wide energy range, 
from 100A MeV to RHIC energies, in a unified way,
it is a big challenge for us to make systematic analyses in these energies.
In this paper, we focus on the mass dependence of the collision
system at AGS energies.
Other applications at higher energies are found elsewhere~\cite{jam1}.

The outline of this paper is as follows.	
We will present a detailed description of cross sections
 and modeling of inelastic processes 
 for $hh$ collisions in section~\ref{sec:hcm},
 because elementary $hh$ processes are essential inputs for the hadronic
 cascade model.
In Sec.~\ref{sec:results}, we first study the transverse momentum
distributions of protons, pions and kaons in
p+Be, p+Al, p+Cu, p+Au, Si+A, Si+Cu and Si+Au collisions
at the laboratory incident momentum of 14.6A GeV/c.
We discuss the role of rescattering by comparing
the cascade model results with the Glauber type calculations.
We then discuss the collision dynamics
 for truly heavy ion colliding system Au+Au collisions.
The summary and outlook are given in Sec.~\ref{sec:summary}.

\section{MODEL DESCRIPTION}\label{sec:hcm}

In this section, we present the assumptions
and parameters of our model together with the inclusive and the exclusive
$hh$ data including incident energy dependence.

\subsection{MAIN COMPONENTS OF THE MODEL}\label{subsec:comp}

The main components of our model are as follows.
(1) Nuclear collision is assumed to be described by the
    sum of independent binary $hh$ collisions.
    Each $hh$ collision is realized by the closest distance approach.
    In this work, no mean field is included, therefore 
    the trajectory of each hadron is straight in between two-body
    collisions, decays or absorptions.
(2) The initial position of each nucleon is sampled by the
    parameterized distribution of nuclear density.
    Fermi motion of nucleons are assigned
    according to the local Fermi momentum.
(3) All established hadronic states, including resonances,
    are explicitly included
    with explicit isospin states as well as their anti-particles.
    All of them can propagate in space-time.
(4) The inelastic $hh$ collisions produce resonances at low energies
    while at high energies
    ( $\gsim 4\GeV$ in $BB$ collisions $\gsim 3\GeV$ in $MB$ collisions
    and $\gsim 2\GeV$ in $MM$ collisions)
    color strings are formed and they decay into hadrons according to the
    Lund string model~\cite{pythia}.
    Formation time is assigned to hadrons from string fragmentation.
    Formation point and time are determined by assuming yo-yo formation point.
    This choice gives the formation time of
    roughly 1 fm/c with string tension $\kappa=1 $GeV/fm.
(5) Hadrons which have original constituent quarks can scatter
    with other hadrons assuming the additive quark cross section
    within a formation time.
    The importance of this quark(diquark)-hadron interaction
    for the description of baryon stopping
    at CERN/SPS energies was reported by Frankfurt group~\cite{rqmd1,urqmd}.
(6) Pauli-blocking for the final nucleons in two-body collisions
    are also considered.
(7) We do not include any medium effect
    such as string fusion to rope~\cite{venus,rqmd1},
    medium modified cross sections and in-medium mass shift.
    All results which will be presented in this paper
    are those obtained from the free cross sections and free masses as inputs.

%
%
%
\subsection{Baryon-baryon interactions}

Let us start with the explanation of the
  resonance model for baryon-baryon ($BB$) collisions
  implemented in our model in detail.
We assume that inelastic $BB$ collisions are described by
	the resonance formations and their decays below C.M. energy
	$\srt<4$GeV
     and at higher colliding energies,
         string formation and their fragmentation into hadrons
         are included based on a similar picture to that
         in the RQMD~\cite{rqmd2} and the UrQMD model~\cite{urqmd}.
The total and elastic $pp$ and $pn$ cross sections are well known.
Fitted cross sections and experimental data
 are shown in Fig.~\ref{fig:nntotal}.
Inelastic cross sections are assumed to be filled up with
the resonance formations up to $\srt=$3-4GeV.
At higher energies, 
the difference between experimental inelastic cross section
 and resonance formation cross sections
 are assigned to the string formation.
\FIGnntotal

The following non-strange baryonic resonance excitation channels
 are implemented for the nucleon-nucleon scattering in our model:
\begin{itemize}
\item[(1)] $NN \to  N\Delta(1232)$,
  \ \ (2) $NN \to  NN^* $,
  \ \ (3) $NN \to  \Delta(1232)\Delta(1232)$,
\item[(4)] $NN \to  N\Delta^* $,
  \ \ (5) $NN \to  N^*\Delta(1232) $,
\ \ (6) $NN \to  \Delta(1232)\Delta^* $,
\ \ (7) $NN \to  N^*N^*  $,
\item[(8)] $NN \to  N^*\Delta^* $,
\ \ (9) $NN \to  \Delta^*\Delta^* $.
\end{itemize}
Here $N^*$ and $\Delta^*$ represent higher baryonic states below 2 GeV/$c^2$.
The $pp$ and $pn$ cross sections are
 calculated from each isospin components $\sigma(I)$
(in some cases we ignore the interferences between different amplitudes):
\begin{equation}
 \sigma(h_1 h_2 \to h_3 h_4) = \sum_I |C(h_1h_2,I)|^2|C(h_3 h_4,I)|^2 \sigma(I)
  \label{eq:crossres}
\end{equation}
where $C(h_i h_j,I)$ is isospin Clebsch-Gordon coefficients.
%
%
For $N^*$ and $\Delta^*$ productions,
the sum of production cross sections of several resonance species
($N(1440) \sim N(1990)$ for $N^*$
	and $\Delta(1600) \sim \Delta(1950)$ for $\Delta^*$)
are parameterized, and resonance species are chosen afterward (see below).
The strength of each branches $\sigma(I)$ are determined from the exclusive
pion production data~\cite{CernHera}.
Isospin $I=1$ component for $NN$ collisions can be extracted
from the $pp$ reactions.
We assume that isospin $I=0$ components are determined from the $pn$ reactions,
then explicit form of cross sections in different isospin channels 
can be written down as follows,
\begin{eqnarray}
\sigma(pp\to p\Delta^+) &=& {1\over4}\sigma(I=1), \qquad
                         \sigma(pn\to n\Delta^+) = {1\over4}\sigma(I=1), \\
\sigma(pp\to n\Delta^{++}) &=& {3\over4}\sigma(I=1),\qquad
                        \sigma(pn\to p\Delta^0) = {1\over4}\sigma(I=1), \\
\sigma(pp\to pp^*) &=& \sigma(I=1) , \qquad
         \sigma(pn\to np^*) = {1\over4}\sigma(I=1)+{1\over4}\sigma(I=0), \\
\sigma(pp\to \Delta^+\Delta^+) &=& {2\over5}\sigma(I=1) , \qquad
 \sigma(pn\to \Delta^0\Delta^+) 
	= {1\over20}\sigma(I=1)+{1\over4}\sigma(I=0),\\
\sigma(pp\to \Delta^0\Delta^{++}) &=& {3\over5}\sigma(I=1) ,\qquad
 \label{eq:dda}
 \sigma(pn\to \Delta^-\Delta^{++})
	={9\over20}\sigma(I=1)+{1\over4}\sigma(I=0)\ .
 \label{eq:ddb}
\end{eqnarray}
The functional form for the non-strange baryonic resonance formation cross
 sections is assumed to be
\begin{equation}
  \sigma(\srt) =
   {{\rm a}(\srt / \srt_{th}-1)^{\rm b} {\rm d}
                 \over (\srt/{\rm c}-1)^2 + {\rm d}^2} ~.
\end{equation}
All parameters
  except one-$\Delta$ production cross section
  are listed in tables~\ref{table:bbresa} and ~\ref{table:bbresb}
  for each isospin channel
  where all cross sections are given in mb and $\srt_{th}$ denotes a threshold.
One-$\Delta$ production cross section
   $\sigma(NN \to N \Delta(1232))$ is
   parameterized with the following functional form
\begin{equation}
\sigma_1(NN\to N\Delta(1232))=
{0.0052840\sqrt{\srt/2.0139999-1}\over (\srt-  2.11477)^2+0.0171405^2}
+{  28.0401(\srt/ 2.124-1)^{ 0.480085} \over((\srt/  2.06672)-1)^2+ 0.576422^2},
\end{equation}
in order to ensure correct threshold behavior.
Pionic fusion cross section ($pp\to d\pi^+$) has been fitted as
\begin{equation}
\sigma(pp\to d \pi^+)=
{  0.14648(\srt/ 2.024-1)^{  0.20807} \over((\srt/  2.13072)-1)^2+ 0.042475^2}
+{  0.12892(\srt/ 2.054-1)^{  0.08448} \over((\srt/  2.18138)-1)^2+ 0.059207^2}.
\end{equation}
In actual simulations,
we have effectively included the cross section of the $NN\to \pi d$
  into the $\Delta$ production cross section for simplicity.
Similarly the s-wave pion production channels $NN\to NN\pi_{s}$
  are simulated as the $N(1440)$ production.
%
%
%
%
\begin{table}[]
\caption[NNND]{Resonances cross section parameters for I=1,
$\pi_s$ denotes s-wave pion production.}
\label{table:bbresa}
\begin{center}
\begin{tabular}{lccccc}
Channel          & a  & b & c & d & $\srt_{th}$ \\\hline
$\sigma_1(NN\to NN^*)$                         &  24.94700 &   2.48150 &   2.63330 &  0.425358 &  2.162\\
$\sigma_1(NN\to \Delta(1232)\Delta(1232))$     &   7.63181 &   1.41140 &   2.67784 &  0.311722 &  2.252\\
$\sigma_1(NN\to N\Delta^*)$                    &   8.01615 &   2.74161 &   3.34503 &  0.259703 &  2.340\\
$\sigma_1(NN\to N^*\Delta(1232))$              &  13.14580 &   2.06775 &   2.75682 &  0.247810 &  2.300\\
$\sigma_1(NN\to \Delta(1232)\Delta^*)$         &  19.63220 &   2.01946 &   2.80619 &  0.297073 &  2.528\\
$\sigma_1(NN\to N^*N^*)$                       &  11.67320 &   2.31682 &   2.96359 &  0.259223 &  2.438\\
$\sigma_1(NN\to N^*\Delta^*)$                  &   2.99086 &   2.29380 &   3.54392 &  0.090438 &  2.666\\
$\sigma_1(NN\to \Delta^*\Delta^*)$             &  35.13780 &   2.25498 &   3.14299 &  0.215611 &  2.804\\
$\sigma_1(NN\to NN\pi_s)$ & 15.644100 &  1.675220 &   2.07706 &  0.658047 &  2.014\\                     
\end{tabular}
\end{center}
\end{table}

\begin{table}[hbtp]
\caption[]{Resonances cross section parameters for I=0,
$\pi_s$ denotes s-wave pion production.}
\label{table:bbresb}
\begin{center}
\begin{tabular}{lccccc}
Channel          & a  & b & c & d & $\srt_{th}$ \\\hline
$\sigma_0(NN\to NN^*)$                         & 166.60600 &   2.10128 &   2.34635 &  0.284955 &  2.162\\
$\sigma_0(NN\to \Delta(1232)\Delta(1232))$     &  39.99770 &   1.83576 &   2.40348 &  0.288931 &  2.252\\
$\sigma_0(NN\to \Delta(1232)\Delta^*)$         &  56.32490 &   2.00679 &   2.71312 &  0.362132 &  2.528\\
$\sigma_0(NN\to N^*N^*)$                       &   2.14575 &   0.21662 &   3.40108 &  0.252889 &  2.438\\
$\sigma_0(NN\to \Delta^*\Delta^*)$             &   4.14197 &   1.67026 &   3.75133 &  0.476595 &  2.804\\
$\sigma_0(NN\to NN\pi_s)$ & 78.868103 &  0.746742 &   1.25223 &  0.404072 &  2.014\\                     
\end{tabular}
\end{center}
\end{table}

\FIGnninel
In Fig.~\ref{fig:nninel},
We show the contributions of non-strange baryonic resonance cross sections
 for different partial channels as functions of c.m. energies.
The upper panels of Fig.~\ref{fig:nninel}
 show the one-resonance production cross section $NN\to NR$ (solid lines),
 two-resonance production cross section $NN\to RR$ (dotted lines),
 the sum of $NR$ and $RR$ cross section (long dashed lines)
 for $pp$ (left panels) and $pn$ (right panels) reactions.
Total inelastic cross sections are filled up by the resonance productions
up to about $E_{cm}=$4GeV, while at CERN/SPS energies, string excitation
is dominated.
The dot-dashed lines in the upper panels of Fig.~\ref{fig:nninel}
   express the string excitation cross sections for $pp$ and $pn$.
At AGS energies corresponding to the
   invariant mass \srt$\sim 5$GeV,
   the contributions of the resonance productions and string productions
   are approximately the same in the first nucleon-nucleon collision
   in our parameterization.
The collision spectrum in $BB$ collisions, however, are spread in broad
  energy range for Au+Au collision as shown in Ref.~\cite{urqmd},
  due to the high baryon density achieved at AGS energies.
Low energy cross sections, therefore, is also important
  in order to treat the dynamics correctly at AGS energies from first chance
  $NN$ collisions to the final hadronic gas stage.

The cross section for the resonance productions may be written by
\begin{equation}
 \label{eq:cross}
 {d\sigma_{12\to34}\over d\Omega}=
    {(2S_3+1)(2S_4+1)\over 64\pi^2 s p_{12}}
    \int\int p_{34} |\Mx|^2 A(m_3^2)A(m_4^2)d(m^2_3)d(m^2_4) ~,
\end{equation}
where $S_i, i=3,4$ express the spin of the particles in the final state.
Mass distribution function $A(m_i)$ for nucleons is just a $\delta$-function,
while that for resonances is given
by the relativistic Breit-Wigner function
\begin{equation}
  A(m^2)={1\over \cal N}{m_R\Gamma(m)\over (m^2-m_R^2)^2 + m_R^2\Gamma(m)^2}.
 \label{eq:rlorentz}
\end{equation}
where $\cal N$ denotes the normalization constant.
 In this paper,
we use simply take ${\cal N} = \pi$ which is a value in the case of
a constant width.
The full width $\Gamma(m)$ is a sum of all partial decay width
$\Gamma_R(MB)$ for resonance $R$ into mesons $M$ and baryons $B$
which depends on the momentum of the decaying particle~\cite{rqmd2,urqmd}:
\begin{equation}
 \Gamma_R(MB)=\Gamma^0_R(MB) {m_R\over m}
              \left({p_{cms}(m)\over p_{cms}(m_R)}\right)^{2\ell+1}
   {1.2 \over 1+0.2\left({p_{cms}(m)\over p_{cms}(m_R)}\right)^{2\ell+1}}
   \label{eq:width}
\end{equation}
where $\ell$ and $p_{cms}(m)$ are
 the relative angular momentum
 and 
 the relative momentum in the exit channel in their rest frame.

The Monte Carlo procedure is as follows.
First, final resonance types $\Delta(1232)$, \Ns or \Ds are chosen 
  using parameterized cross sections
 and then
 we determine each resonance production channel
  according to the equation Eq.(\ref{eq:cross}).
To do this, we need to know the matrix element $|\Mx|^2$ for all resonances.
In the present model, we make a simple assumption that
 each resonance production cross sections can be selected
 according to the probability:
\begin{equation}
  \label{eq:resprob}
  P(R_i,R_j) \sim (2S_i+1)(2S_j+1)\int\int 
                    p_{ij}A_i(m^2_i)A_j(m^2_j)d(m^2_i)d(m^2_j)~.
\end{equation}
Namely, the partial cross section for each resonance state
is only governed by the final spins and mass integrals
ignoring the resonance state dependence of the matrix element.
Once types of resonances are chosen,
 we generate the resonance masses according to the distribution
  neglecting the mass dependence of the matrix elements in Eq.~(\ref{eq:cross})
  \begin{equation}
    P(m_3,m_4)dm_3dm_4 \sim 4m_3m_4p_{34}A(m^2_3)A(m^2_4)dm_3dm_4 \ ,
  \end{equation}
in the reaction $1+2 \to 3+4$,  where mass distribution function $A(m_i)$
  should be replaced by the $\delta$-function
  in the case of stable particles in the final state.

In the case of the collisions involving resonance states
 in the ingoing channel,
we use the approximation that the inelastic cross sections
for resonance productions
 as well as
 the elastic cross section are the same as the nucleon-nucleon
cross sections at the same momentum in the c.m. frame,
except for de-excitation processes, $NR \to NN$ and $RR \to NN$,
whose cross sections are estimated by using the detailed balance
described in the next subsection, \ref{subsec:detbal}.

\FIGnnex
Fig.~\ref{fig:nnex} shows 
 the energy dependence of the exclusive pion production cross sections
 in $pp$ (up to five pion production) and $pn$ (up to two pion production)
 reactions.
We compare the results obtained from our simulation
  with the data~\cite{CernHera}.
Overall agreement is achieved in these exclusive pion productions
within a factor of two with the above simplification of
 the common matrix element in Eq.~(\ref{eq:resprob}).
Smooth transition from the resonance picture to the string picture
at $E_{cm}= 3\sim4$ is
achieved since no irregularity of the energy dependence is
present in the calculated results.
String excitation law will be described later in section~\ref{sec:string}.
In order to get more satisfactory fit, for example, we can improve the model
by introducing different values for the matrix elements
for different resonance channels.
For example, in Ref.~\cite{Teis}, the matrix elements are fitted to reproduce
the pion production cross sections up to two-pion productions
 as well as $\eta$ production cross section
  assuming that they are independent of masses but dependent on species.

\subsection{Detailed balance}
\label{subsec:detbal}

In the processes of resonance absorption,
  we use a generalized detailed balance formula~\cite{detbal1,detbal2,Wolf2}
  which takes the finite width of the resonance mass into account.
The time-reversal invariance of the matrix element leads to the
principle of detailed balance. If scattering particles are all stable,
the formula is given by
\begin{equation}
  {d\sigma_{34\to 12}\over d\Omega}=
       {(2S_1+1)(2S_2+1)\over (2S_3+1)(2S_4+1)}
          {p^2_{12} \over p^2_{34}}
    {d\sigma_{12\to 34}\over d\Omega}
\end{equation}
where $S_i$ denotes the spin of particle $i$ and $p_{ij}$
corresponds to the c.m. momentum of the particles $i$ and $j$.

The differential cross section for the reaction $(1,2)\to(3,4)$
for the stable particles may be written by
\begin{equation}
  {d\sigma_{12\to 34}\over d\Omega}=
       {|\Mx_{12\to34}|^2 \over 64\pi^2 s}
       {1\over (2S_1+1)(2S_2+1)}
       {1 \over p_{12}}
	\int \int\ p_{34}\,
       \delta(p_3^2-m_3^2)d(m_3^2)
       \delta(p_4^2-m_4^2)d(m_4^2)
\end{equation}
where $|\Mx_{12\to34}|^2$ represents the spin-averaged matrix element.
If the particles have a finite width, we should replace above $\delta$
functions to the certain normalized  mass distribution functions $A(m)$.
Using the $|\Mx_{12\to34}|=|\Mx_{34\to12}|$,
we obtain
\begin{equation}
  {d\sigma_{34\to 12}\over d\Omega}=
       {(2S_1+1)(2S_2+1)\over (2S_3+1)(2S_4+1)}
          {p^2_{12} \over p_{34}}
    {d\sigma_{12\to 34}\over d\Omega}
     {1\over \int\int p_{34} A(m_3^2)A(m_4^2)d(m_3^2)d(m_4^2)}\ ,
\end{equation}
where we use the relativistic Breit-Wigner function
 Eq.(\ref{eq:rlorentz}) for mass distribution function $A(m^2)$.
The extra factor compared to the usual detailed balance formula
  increases  the absorption cross section.
It has been proved that
   this formula plays an essential role
   in order to understand
   the $\pi N \Delta$ dynamics~\cite{detbal1,detbal2,Wolf2}.

In the case of one-\Dl~ absorption cross section,
   we can write down the following formula:
\begin{equation}
\sigma_{N\Delta\to NN'}={1\over2}
        {1\over 1+\delta_{NN'}}
     {p^2_N\over p_{\Delta}}
     \sigma_{NN' \to N\Delta}
     \left(
       \int^{(\srt-m_N)^2}_{(m_N+m_{\pi})^2} p_{\Dl}A(m^2)dm^2
     \right)^{-1} \label{delbal} \ .
\label{eq:detbaldelta}
\end{equation}
$p_N$ and $p_{\Delta}$ are
    the final nucleon-nucleon  c.m. momentum
    and the initial c.m. momentum, respectively.
The factor $1/(1+\delta_{NN'})$ in Eq.(\ref{delbal})
  arises from the identical nature of the final states,
 and $1/2$ comes from spins.

There are some versions of the extended detailed balance formula,
which are slightly different from each other.
For example, 
Danielewicz and Bertch~\cite{detbal1} use the formula
\begin{equation}
  {d\sigma_{34\to 12}\over d\Omega}=
       {(2S_1+1)(2S_2+1)\over (2S_3+1)(2S_4+1)}
          {p^2_{12} \over p_{34}}
          {m_3 \over m^R_3}
          {m_4 \over m^R_4}
    {d\sigma_{12\to 34}\over d\Omega}
     {1\over \int\int p_{34} A(m^{'2}_3)A(m^{'2}_4)dm_3^{'2}dm_4^{'2}} ,
\end{equation}
here $m_i^R$ denotes the pole mass of the resonance $i$,
while Wolf, Cassing and Mosel~\cite{Wolf2} use
\begin{equation}
  {d\sigma_{34\to 12}\over d\Omega}=
       {(2S_1+1)(2S_2+1)\over (2S_3+1)(2S_4+1)}
          {p^2_{12} \over p^2_{34}}
    {d\sigma_{12\to 34}\over d\Omega}
     {1\over \int\int A(m_3^2)A(m_4^2)dm_3^2dm_4^2} .
\end{equation}
We have checked that these formulae give similar results to ours.
Fig.~\ref{fig:detbal} shows the comparisons between the different
 formulae of the cross sections
for the reaction $\Delta^{++}n\to pp$.
\FIGdetbal

\subsection{Meson-Baryon, Meson-Meson Collisions}
\label{subsec:MBMM}

We now turn to the explanation of meson-baryon ($MB$) and meson-meson
($MM$) collisions.
We also use
 resonance/string excitation model for $MB$ and $MM$ collisions.

\FIGxpin
Total cross section for $\pi N$ ingoing channel is assumed to be decomposed to
\begin{equation}
 \sigtot^{\pi N}=\sigbw+\sigel +\sigsS+\sigtS,
\end{equation}
where \sigel, \sigbw, \sigsS, and \sigtS~ denote
the $t$-channel elastic cross section,
the $s$-channel resonance formation cross section
	with the Breit-Wigner form, 
the $s$-channel and $t$-channel string formation cross sections, respectively.
We neglect the $t$-channel resonance formation cross section at a
  energy range of 
$\sqrt{s}\lsim 2$GeV.
The $t$-channel elastic cross section \sigel~ was determined so that
the sum of the elastic component of
 the $s$-channel Breit-Wigner cross section \sigbw~
 and $t$-channel elastic cross section \sigel~ 
reproduces the experimental elastic data for $\pi N$ interaction.
Above the $\Delta(1232)$ region, $t$-channel elastic cross section
becomes non-zero in our parameterization 
(Figs.~\ref{fig:xpip} and \ref{fig:xpin}).
String formation cross sections (\sigsS~ and \sigtS) are 
 determined to fill up the difference between
 experimental total cross section and $\sigbw+\sigel$.
We calculate the resonance formation cross section \sigbw\ 
using the Breit-Wigner formula~\cite{Brown,rqmd1}
(neglecting the interference between resonances),
\begin{equation}
    \label{eq:bw}
\sigma(MB \to R)= {\pi(\hbar c)^2 \over p_{cm}^2}
         \sum_{R}  |C(MB,R)|^2  {(2S_R+1) \over (2S_M+1)(2S_B+1)}
                     {\Gamma_R(MB)\Gamma_R(tot) \over
                      (\sqrt{s}-m_R)^2+\Gamma_R(tot)^2/4} \ .
\end{equation}
The momentum dependent decay width
Eq.~(\ref{eq:width}) are also used for the calculation of
 decay width in Eq.~(\ref{eq:bw}).
$S_R$, $S_B$ and $S_M$ denote the spin of
 the resonance, the decaying baryon and meson respectively.
The sum runs over resonances,
   $R=N(1440)\sim N(1990)$ and $\Delta(1232)\sim\Delta(1950)$.
Actual values for these parameters are 
 taken from the Particle Data Group~\cite{PDG96}
 and adjusted
 within an experimental error bar
to get reasonable fit for $MB$ cross sections.
The results of the fit are shown in Figs.~\ref{fig:xpip},\ref{fig:xpin}.
It has been shown that
  inclusion of resonances
  play an important role to study strangeness
  productions in AGS and SPS energies~\cite{rqmd1}
  and $(K^-,K^+)$ reactions~\cite{NOHE97}.
In fact,
 strangeness production cross sections
 for resonance-$N$ ingoing channels
  are found to be much larger
  than that for $\pi N$ channel.
This would be effective to explain
the strangeness enhancement observed in heavy ion collisions
within a rescattering scenario.

\FIGxakn
Since the $\Kbar N$ interaction
has some exoergic channels 
such as $\Kbar N\to \pi Y$,
we need to include additional terms:
\begin{equation}
 \sigtot^{\Kbar N}=\sigbw+\sigel+\sigch+\sigpiY+\sigsS+\sigtS,
\label{eq:xakn}
\end{equation}
where \sigch~ and \sigpiY~ denote $t$-channel charge exchange reaction
  and $t$-channel hyperon production cross sections
  which are also fixed by the requirement that
  the sum of $t$-channel contributions and
  Breit-Wigner contributions reproduce experimental data.
Breit-Wigner formula enables us to calculate
  experimentally unmeasured cross sections
  such as $\rho N \to \Lambda K$.
For the calculation of \sigbw,
we include hyperon resonances,
      $R=\Lambda(1405)\sim\Lambda(2110)$
      and 
      $\Sigma(1385)\sim\Sigma(2030)$.
The total and elastic cross sections for $\Kbar N$ interactions
 used in JAM are plotted in Figs.~\ref{fig:akn1} and \ref{fig:akn2}
 in comparison with experimental data~\cite{PDG96}.

\FIGxkny
The symbol $\sigma_{\pi Y}(s)$ in Eq.(\ref{eq:xakn})
 is the sum of $t$-channel pion hyperon production
cross sections $\Kbar N\to \pi Y$, $Y=\Lambda,\Sigma$.
In Fig.~\ref{fig:xkny},
   we plot the cross sections of
   hyperon productions and charge exchange cross sections
   as well as Breit-Wigner contributions
   fitted in our model.
The cross section for the inverse processes such as $\pi Y\to  \Kbar N$ are
calculated using the detailed balance formula.

\FIGxkaon
$KN$ ingoing channel cannot form any $s$-channel resonance
 due to their quark contents. Therefore
the total cross section can be written within our model as follows, 
\begin{equation}
 \sigtot^{KN}=\sigtR+\sigel+\sigch +\sigtS,
\end{equation}
where \sigtR~ is $t$-channel resonance formation cross section.
Total, elastic and charge exchange cross sections used in our model
are shown in Fig.~\ref{fig:xkaon}.
In the present version of JAM,
only $K N\to K\Delta$, $K N\to K(892)N$ and $K N \to K(892)\Delta$
are explicitly fitted to experimental data~\cite{CernHera}
and fitted results are shown in Fig.~\ref{fig:xkaonD}.

\FIGxkaonD

In meson-meson scattering, we also 
apply the same picture
as that in meson-baryon collisions:
\begin{equation}
 \sigtot=\sigbw + \sigtR + \sigel + \sigsS + \sigtS.
\end{equation}
The difference between experimental inelastic cross section
and resonance cross sections at energies above resonance region
for the meson-baryon and meson-meson collisions
are attributed to the string formation cross section
where $1/\srt$ energy dependence of \sigsS~ is used~\cite{rqmd2}.

For the cross sections for which no experimental data are available,
we calculate the total and elastic cross sections by using the
additive quark model~\cite{goulianos,rqmd1,urqmd}.
\begin{eqnarray}
  \sigma_{tot} &=& \sigma_{NN}\left({n_{1}\over3}\right)\left({n_2\over3}\right)
                 \left(1-0.4{n_{s1}\over n_1}\right)
                 \left(1-0.4{n_{s2}\over n_2}\right) ~, \\
  \sigma_{el} &=& \sigma_{tot}^{2/3}
               (\sigma_{el}^{NN}/ \sigma_{NN}^{2/3}) ~,
\end{eqnarray}
where $\sigma_{NN}$, $\sigma_{el}^{NN}$ express nucleon-nucleon total
and elastic cross sections and $n_i$,$n_{si}$ are the 
number of quarks and $s$-quarks contained in the hadron respectively.
This expression works well above resonance region where the cross section
becomes flat.

For the $t$-channel resonance production cross sections \sigtR,
we do not fit experimental data explicitly in this work
except for $NN$ reaction and one and two pion productions in $KN$ reaction,
because of the vast body of the possibilities for the final states.
 Instead, we simply determine the outgoing resonance types
according to the spins $S_3$, $S_4$ in the final state  and phase space
for the production of resonances $R_3$ and $R_4$
\begin{equation}
 P(R_3,R_4) \propto (2S_3+1)(2S_4+1)p_{34}(s)^2 ~.
\end{equation}
where $p_{34}(s)$ denotes the c.m. momentum in the final state.
If the ingoing channel involves resonances, their ground state
particles are also considered in the final state.
Once the outgoing resonance types are determined,
we generate masses according to the Breit-Wigner distribution.

For the angular dependence in the processes of
  $t$-channel resonance production \sigtR,
we use
\begin{equation}
  {d\sigtR \over dt} \sim \exp(bt),
\end{equation}
and the slope parameter $b$ for the energy range of
  $\srt>2.17\GeV$
   is parameterized by
\begin{equation}
   b=2.5 + 0.7\log(s/2),
\end{equation}
with invariant mass squared $s$ given in units of GeV$^2$.
We use the same parameterization presented in Ref.~\cite{niita}
 for the energy below $\srt<2.17\GeV$ for the $t$-channel resonance
productions.
The elastic angular distribution is also taken from Ref.~\cite{niita}
 for $\srt<10\GeV$
and from PYTHIA~\cite{pythia} for $\srt>10\GeV$.

\subsection{String formation and fragmentation} \label{sec:string}

At an energy range above $\srt>4-5$GeV,
the (isolated) resonance picture breaks down because
width of the resonance becomes wider and 
the discrete levels get closer.
The hadronic interactions at
the energy range 4-5$<\srt<$10-100GeV where
it is characterized by the small transverse momentum transfer
is called "soft process", and
string phenomenological models
are known to describe the data for such soft interaction well.
The hadron-hadron collision 
leads to a string like excitation longitudinally.
In actual description of the soft processes,
we follow the prescription adopted in the HIJING model~\cite{hijing}, 
as described below.

In the center of mass frame of two colliding hadrons,
we introduce light-cone momenta defined by
\begin{equation}
p^+ = E+p_z, \qquad p^- = E-p_z\ .
\end{equation}
Assuming that beam hadron 1 moves in the positive z-direction
and target hadron 2 moves negative z-direction,
the initial momenta of the both hadrons are
\begin{equation}
  p_1 = (p_1^+,p_1^-,0_T), \qquad p_2 = (p_2^+,p_2^-,0_T)\ . 
\end{equation}
After exchanging the momentum $(q^+,q^-,\pp_T)$, the momenta will change to
\begin{equation}
       p'_1 = ((1-x^+)P^+,x^-P^-,\pp_T),
 \qquad p'_2 = (x^+P^+,(1-x^-)P^-,-\pp_T), 
\end{equation}
where $P^+=p_1^++p_2^+=P^-=p_1^-+p_2^-=\srt$ (in c.m. frame).
The string masses will be
\begin{equation}
  M_1^2= x^-(1-x^+)s-p^2_T, \qquad M_2^2=x^+(1-x^-)s-p^2_T,
\end{equation}
respectively.
Minimum momentum fractions are
 $x_{min}^+=p_2^+/P^+$ and $x_{min}^-=p_1^-/P^-$.
For light-cone momentum transfer for the non-diffractive events,
 we use the same distribution as that in
DPM~\cite{dpm} and HIJING~\cite{hijing}:
\begin{equation}
  P(x^{\pm})= {(1.0-x^{\pm})^{1.5}\over (x^{\pm2}+c^2/s)^{1/4}}\ ,
\end{equation}
for baryons and
\begin{equation}
  P(x^{\pm})= {1\over (x^{\pm2}+c^2/s)^{1/4}((1-x^{\pm})^2+c^2/s)^{1/4}}\ ,
\end{equation}
for mesons, where $c=0.1 $GeV is a cutoff.
For single-diffractive events, in order to reproduce
experimentally observed mass distribution $dM^2/M^2$,
we use the distribution
\begin{equation}
   P(x^{\pm})={1 \over (x^{\pm2}+c^2/s)^{1/2}}.
\end{equation}

The strings are assumed to hadronize via quark-antiquark 
or diquark-antidiquark 
creation using Lund fragmentation model PYTHIA6.1\cite{pythia}.
Hadron formation points from a string fragmentation
are assumed to be given by
  the yo-yo formation point~\cite{bialas} which is defined by the 
  first meeting point of created quarks.
Yo-yo formation time is about 1fm/c assuming the string
tension $\kappa=1$ GeV/fm at AGS energies.

The same functional form as the HIJING model~\cite{hijing}
  for the soft $\pp_T$ transfer at low 
$p_T<p_0$
is used
\begin{equation}
   f(\pp_T) = \left\{ (p_T^2+c_1^2)(p_T^2+p_0^2)
                     (1+e^{(p_T-p_0)/c_2})  \right\}^{-1} ~,
\end{equation}
where $c_1=0.1\GeVc$, $p_0=1.4\GeVc$ and $c_2=0.4\GeVc$,
to reproduce the high momentum tail of the particles
at energies $E_{lab}=10\sim20$GeV.

\FIGppxa
In Fig.~\ref{fig:ppxa}, the calculated rapidity distributions of
protons, positive and  negative pions for proton-proton collisions
 at 12GeV/c and 24GeV/c are shown with the data from Ref.\cite{pp1224exp}.
The proton stopping behavior and the pion yields
are well described by the present model.
Within our model,
 fast protons come from resonance decays
  and
  mid-rapidity protons from string fragmentation of
Lund model (PYTHIA6.1~\cite{pythia}) with the default parameter
which determines the probability
 of diquark breaking at these energies.
Anisotropic angular distribution in a resonance decay is
taken into account assuming Gaussian $p_T$ distribution ~\cite{rqmd2}
with a mean value of 0.35GeV/c$^2$.

\FIGppxb

As reported in HIJING~\cite{hijing},
an extra low $p_{T}$ transfer to the constituent
quarks is important to account for
the high $p_{T}$ tails
of the pion and proton distributions at energies $\plab \sim 20$GeV/c$^2$.
As shown in Fig.~\ref{fig:ppxb},
the present model also reproduces 
the proton and pion transverse momentum distributions reasonably well
at \plab=12GeV/c$^2$ and 24 GeV/c$^2$.

\FIGppx
In addition, the present model also describes well the energy dependence
of the particle production cross sections, as shown in Fig.~\ref{fig:ppx}.
Here we show the incident energy dependence of
 the inclusive data
for $\pi^+$, $\pi^-$, $\pi^0$, $K^0$,$K^+$,$\Lambda$, $\Sigma^-$
and $\Sigma^0$ productions
 from proton-proton interactions in comparison with the experimental
data~\cite{CernHera}.

The comparisons shown until now in Figs.%
2 -- 13
show that
the combination of particle production mechanisms by the resonance decay
and the string decay enables us to explore a wide incident energy region
from a few hundred MeV to a few ten GeV,
with reasonably well fitted inclusive as well as exclusive cross sections,
which are essential inputs in cascade models.

\section{RESULTS}\label{sec:results}

In the following,
we systematically apply our hadronic cascade model (JAM1.0~\cite{jam})
to proton, silicon and gold induced reactions at AGS energies
and investigate the effect of cascading by comparing
the results obtained by the cascade model  with the Glauber
type calculations.

\subsection{COMPARISON TO E802 DATA}

In this section,
 we first focus our attention on the proton transverse distributions
to check the detailed examination of the collision term
and its space-time picture (formation time) used in our model
 and also to see the transition of the reaction dynamics from proton induced
collisions to heavy ion collisions.

\FIGproton
We show in Fig.~\ref{fig:proton}
 proton invariant transverse mass distributions
calculated by our cascade model
for the proton induced reactions, p+Be, p+Al, p+Cu, p+Au,
and silicon induced reactions, central Si+Al, Si+Cu, Si+Au
at 14.6GeV/c in comparison to the data from the E802
collaboration~\cite{E802pA,E802b}.
In each figure, spectra are plotted in a rapidity interval of 0.2
and are displayed by multiplying each by a power of 10 
from bottom to upper.
Si+Al, Si+Cu and Si+Au data correspond to the central collision
 with 7 \% centrality. 
For the calculations of Si+A (A=Al,Cu,Au) systems,
impact parameter are distributed $b<1.797\fm$ for Si+Al,
$b<2.2\fm$ for Si+Cu and $b<2.9\fm$ for Si+Au.
Our calculations show good agreement with the data in proton induced reactions.
In silicon induced reactions,
 our calculations well account for the experimental data in general.
However, we can see some overestimate 
at low transverse momenta, in particular for Si+Au system.
As a result, our cascade model gives larger proton stopping than the data.

\FIGprotong
Now we compare the cascade model results with the Glauber type calculation
in order to see the effect of pion rescattering, nucleon cascading, 
and the consequent deviation from the linear extrapolation
of sum of binary nucleon-nucleon collisions to proton-nucleus and
nucleus-nucleus collisions.
The Glauber type models such as the FRITIOF model~\cite{fritiof}
have been widely used at higher energies, i.e. more than 200 AGeV.
We use the same method as the FRITIOF model  with some modifications.
The wounded nucleons
 become resonances or strings in each nucleon-nucleon collisions,
 and strings can interact again before they fragment.
Resonances can be converted to nucleons,
 and strings are allowed to be de-excited to the minimum string mass.
After all binary collisions are completed, 
strings and resonances are forced to decay. 
Rescattering of produced particles are not considered.
In a present treatment, we have used our parameterization 
in calculating the probability to excite nucleons to resonances or strings.
Figure~\ref{fig:protong}
 shows the results obtained by this Glauber type calculation.
In beam rapidity region, 
 for all systems, 
 good agreements can be seen
  because the effect of rescattering would be small as expected.
For proton-induced reactions, 
 the Glauber type calculation gives steeper shape in comparison to the data
 at mid-rapidity and target regions.
In heavy ion reactions, 
 this deviation is significant at around the mid-rapidity.
Rescattering, therefore, is necessary to account for
 transverse momentum distributions 
of protons for reactions involving heavy nuclei.

\FIGpip
\FIGpin
We now turn to the mass dependence of the pion transverse distributions.
In Fig.~\ref{fig:pip} and Fig.~\ref{fig:pin}, we show the calculated
$\pi^+$ and $\pi^-$ spectra by histograms together with
the E802 data~\cite{E802pA,E802b}.
Agreement between the cascade model and the data is very good 
for all the combination of projectile and target,
and for both of the slope parameter and the absolute value
of the cross section.

On the other hand, 
as shown in Fig.~\ref{fig:pipg} and Fig.~\ref{fig:ping}, 
Glauber type calculations 
well reproduce 
the data for small mass systems p+Be, p+Al, p+Cu and Si+Al,
and give similar slopes to the experimental data,
while the multiplicity of pions in heavy systems
 are larger than those of the data. 
This is due to the effect of rescattering in which
pions are absorbed during the evolution for the large mass systems.

\FIGpipg
\FIGping


\FIGkp
\FIGkn
\FIGkpg
\FIGkng

Let us study the kaon and anti-kaon transverse mass spectra of E802.
The calculated transverse mass distributions of $K^+$ and $K^-$
in cascade model
 are compared 
 with the E802 data~\cite{E802pA,E802b}
 in Figs.~\ref{fig:kp} and ~\ref{fig:kn}.
Figures~\ref{fig:kpg} and ~\ref{fig:kng}
  are the results with the Glauber type calculations
  for $K^+$ and $K^-$ invariant transverse momentum distributions.
We find more significant differences 
between the cascade and Glauber results of kaon productions
than those in proton and pion spectra.
This fact shows the importance of the rescattering:
As discussed in \ref{subsec:MBMM}, 
some of the exoergic $MB$ reactions, which involve resonances,
have very large strangeness production cross sections,
and they contribute to $K$ and $\bar{K}$ productions,
especially in heavy ion reactions~\cite{rqmd1,NOHE97}.
Enhancements due to these meson rescattering
are clearly seen in $K^+$ and $K^-$ spectra, 
except for the reactions of p+Be and p+Al,
because there is no meson-baryon collision in Glauber type calculation.

\subsection{Au+Au COLLISIONS}

\FIGAuAumtp

We continue our comparison to E866 experimental data~\cite{E802AuAu} with
the truly heavy ion collision Au+Au at 11.6AGeV.
Our cascade model calculation
 with impact parameter $b\leq 3.338$fm
  is compared to E866 data~\cite{E802AuAu}
  in Fig.~\ref{fig:AuAumtp} from c.m. rapidity of $y=0.05$ to $y=1.05$
  with the rapidity bin of 0.1 scaled down by a factor of 10 from
  the top to  the bottom spectrum.
The cascade model results show good agreement with the data
 at the transverse mass above $0.2$GeV/$c^2$.  
The cascade model, however, overpredicts again
 the proton spectrum in the low transverse momentum region.

\FIGAuAurapmt
In Figs.~\ref{fig:AuAurap} and \ref{fig:AuAumt},
we compare the cascade model results with the experimental data
by E802 collaboration~\cite{E802AuAu}
in central Au+Au collisions.
Pion multiplicities are in good agreement with
data as well as the slopes of both $\pi^+$ and $\pi^-$.
However, the present cascade model does not describe the suppression
 of protons having low transverse momenta,
 and consequently it gives stronger stopping of proton than the data.
This proton rapidity spectrum for central Au+Au collisions
 shows a similar amount of stopping as those with other cascade models
 like RQMD~\cite{rqmd1}, ARC~\cite{arc} and ART~\cite{art}.
In addition, Glauber type calculation gives the same results for 
the proton rapidity distribution.
Therefore, this defect is not a consequence of the cascade model.
Since the deviation of the transverse mass spectrum of heavy hadrons
from a single exponential behavior is generally considered as a result
of the radial flow, it may be influenced by the nuclear mean field.
In fact,
it has been found in the works of RQMD~\cite{mattiello} and ART~\cite{art}
nuclear mean field significantly reduces maximum baryon
densities of the hadronic matter,
 and consequently the midrapidity protons becomes small.

In this work, we have assumed that the elastic and inelastic cross sections
involving baryon resonances, except for the de-excitation processes to $NN$,
are the same as that in $NN$ channel at the same c.m. momentum.
However, since the de-excitation cross sections are enhanced due to
the generalized detailed balance as explained in Section~\ref{subsec:detbal},
if other cross sections is smaller than those of $NN$,
stopping power may be reduced.

\subsection{Mass dependence of the collision dynamics}

In this section, we study the mass dependence of the 
collision dynamics within the hadronic cascade model.
First we present the cascade model results of $BB$, $MB$ and $MM$ collision
 number as a function of system mass ($A+B$) in comparison to that
of Glauber type calculations.

\FIGncoll
Figure~\ref{fig:ncoll} displays the total collision number
of $BB$ (open squares) and $MB$  (open circles) and $MM$ (open crosses)
obtained from cascade calculation 
  together with Glauber type calculation (open diamonds)
  for the p+A (left panel) and Si+A (right panel) collisions.
When system becomes bigger, $BB$ collisions are
much more frequent 
in cascade model than in the Glauber predictions
even in the proton induced collisions.
This indicates that
there are successive nucleon cascading in the nuclear medium
 in the cascade model picture.
It is interesting to see that
  the number of $BB$ and $MB$ collisions are almost
  the same in heavy ion collisions.
This seems to be the origin of 
the pion number suppression, the increase in proton transverse momentum slope,
and the increase in kaon yield.

\FIGmesons
Indeed we can see the reduction of produced pion multiplicity in cascade model
compared to the Glauber type calculations as shown 
in Fig.~\ref{fig:mesons} where the number of total produced mesons
are plotted as a function of system mass number.
Pions are absorbed mainly in the two-step processes in the cascade model.
For example, the most important pion absorption path
at AGS energies is
\begin{equation}
 \pi N \to \Delta\ , \qquad N\Delta \to NN \ .
\end{equation}
Therefore, large number of $BB$ and $MB$ collisions are necessary
 to describe appropriate pion absorptions.

\FIGcolls
In order to get more detailed information on the mass dependence of the
 collision dynamics at AGS energies,
  we display in Fig.~\ref{fig:colls}
 the colliding energy spectrum of $BB$, $MB$ and $MM$ 
 given by the cascade model together with those 
 from the Glauber typer calculations.
The $BB$ collision distributions as a function of invariant mass
 are very different between the cascade model and the Glauber type model.
The Glauber type calculations predict the collisions
  which are spread around the initial $NN$ c.m. energy,
 while the $BB$ collisions occur at all available collision energies
 in the cascade model.
In both p+A and A+B reactions, 
 the $MB$ collisions are pronounced 
  in the resonance region ($\srt \leq 2$GeV).
It is interesting to note that
  in p+Au system, the number of low energy $BB$ collision 
  is much larger than that of p+Be, p+Al and p+Cu systems.
In A+B systems (bottom of Fig.~\ref{fig:colls}),
 collision number grows very quickly, however,
 the shape of the collision spectrum is similar in all the systems.

\section{SUMMARY}\label{sec:summary}

We have systematically studied the system mass dependence of
the particle distributions at AGS energies with
 a newly developed cascade model (JAM1.0).
The cascade model is shown to provide
a good description of the observed data
for various combination of projectile and target  
without any change of model parameters.
The effect of rescattering of produced particles and
nucleon cascading are found to be important
to explain both the pion yield and the transverse slope,
which are demonstrated by comparing the cascade model results
with the Glauber type calculations.
Those effects increase the transverse momentum slopes of protons and pions,
  and reduce the pion yield.
The importance of the rescattering among particles
 is more visible in kaon spectra.

One of the problems in the hadronic cascade model JAM
is that it gives much larger stopping of the protons
in nucleus-nucleus collisions.
This large stopping is not the consequence of the rescattering
 because Glauber type calculation also gives the same amount of baryon
 stopping.
This problem of strong baryon stopping in cascade models
 has been reported
that proton spectra can be fitted by the inclusion of nuclear mean field
 in RQMD~\cite{mattiello} and ART~\cite{art}.

Another possible solution may be to suppress the cross sections
such as $\sigma(N^*_1 N^*_2 \to N^*_3 N^*_4)$,
which is assumed to be the same as that in $NN$ ingoing channel
at the same c.m. momentum in a present model.
These interactions among resonances become important at AGS energies
 where we have sufficiently dense matter in heavy ion collisions,
and the baryon stopping is sensitive to the cross sections
  in the resonance ingoing reactions.
In fact, we have checked that
if resonance-resonance (BB) cross sections are reduced from nucleon-nucleon
cross section, we get less proton stopping than the present results.
Detailed study in this line will be interesting.

\section*{ACKNOWLEDGMENTS}

One of the author (Y.N.) would like to thank BNL for the kind
 hospitality where a part of the calculations are done
Y. N. also would like to thank Prof. H. St\"ocker and Dr.  H. Sorge
for their encouragements and useful comments.
We acknowledge careful reading of the manuscript by Prof. R. Longacre.


\begin{thebibliography}{9}
\bibitem{jam}JAM1.0, source code and its documentation, to be prepared.
\end{thebibliography}
\end{document}